\documentclass{JHEP3}
\usepackage{amsmath}
\usepackage[mathscr]{eucal}
\newcommand{\bs}{\boldsymbol}
\newcommand{\td}{{\rm d}}
\newcommand{\susy}{\mathbf{Q}}
\newcommand{\brs}{\mathbf{Q}_{\rm B}}
\newcommand{\qtot}{\widehat{\mathbf{Q}}}
\newcommand{\qq}{\mathbf{H}}
\newcommand{\pauli}{\bs{\tau}}

\title{Seiberg-Witten Theories on Ellipsoids}

\author{Naofumi Hama and Kazuo Hosomichi\\
Yukawa Institute for Theoretical Physics, Kyoto University, Japan \\
E-mail:
\email{hama@yukawa.kyoto-u.ac.jp},
\email{hosomiti@yukawa.kyoto-u.ac.jp}
}

\abstract{We present a set of equations for a 4D Killing spinor which
guarantees the Seiberg-Witten theories on a curved background to be
supersymmetric. The equations involve an $SU(2)$ gauge field and some
auxiliary fields in addition to the metric. Four-dimensional ellipsoids
with $U(1)\times U(1)$ isometry are shown to admit a supersymmetry if
these additional fields are chosen appropriately. We compute the
partition function of general Seiberg-Witten theories on ellipsoids, and
the result suggests a correspondence with 2D Liouville or Toda
correlators with general coupling constant $b$.
}

\preprint{YITP-12-51}


\begin{document}

\section{Introduction}\label{sec:intro}

Supersymmetric gauge theories have a characteristic feature that, due to
cancellations of bosonic and fermionic contributions, certain physical
quantities can be evaluated beyond perturbation theory. In this area,  a
number of important exact results have been obtained for the theories
realized on deformed or curved backgrounds which admit rigid
supersymmetry. For example, in 4D ${\cal N}=2$ supersymmetric gauge
theories or Seiberg-Witten (SW) theories, an analytic formula for the
partition function on Omega background was given by Nekrasov in the
pioneering work \cite{Nekrasov}. More recently, an exact formula for
partition function as well as expectation values of Wilson loops on round
four-sphere has been obtained by Pestun \cite{Pestun}. Similar exact results
have also been obtained for 3D ${\cal N}\ge2$ gauge theories on round
three-sphere \cite{Kapustin-WY, Jafferis, Hama-HL1}, its orbifold
\cite{Benini-NY, Alday-FS}, and 2D theories on sphere
\cite{Benini-Cr, Doroud-GLL}. These all served as new powerful tools
to study the strong coupling behavior of the theories
at low energy or other non-perturbative aspects. They also led to a
discovery of a surprising connection between SW theories and 2D
Liouville or Toda conformal field theories, called AGT relation
\cite{AGT,Wyllard}.

So far, most of the work in this field has been focusing on theories on
round spheres. A natural question would then be what other curved
spaces admit rigid supersymmetry. Some systematic analysis has been made
in \cite{Festuccia-S,Jia-S,Samtleben-T,Klare-TZ,Dumitrescu-FS} to draw
conditions on the background geometry from Killing spinor equation. Also,
in \cite{Kallen, Ohta-Y, Kallen-Z} another construction of
supersymmetric gauge theories in three or five dimensions has been
discussed in connection with contact geometry, and moreover some
exact results have been worked out for theories on 3D Seifert
manifolds. On the other hand, there is also a less systematic approach
in which one focuses on a specific class of deformations of round sphere
aiming for a particularly interesting physical consequence.

According to the AGT relation, partition functions of certain class of
SW theories on round $S^4$ agree with correlation functions of 2D
Liouville or Toda Theories at a special value $b=1$ of the coupling.
The coupling $b$ characterizes uniquely the underlying conformal symmetry of
the 2D theory. For example it enters in the Liouville central charge,
\begin{equation}
 c_\text{L}~=~ 1+6Q^2,\quad Q\equiv b+\frac1b.
\end{equation}
One would therefore naturally imagine there is a deformation of
the round sphere which can reproduce the CFT correlators for general
values of the coupling $b$. Actually, similar problem has been resolved
in the setting of a generalized AGT relation involving 3D ${\cal N}=2$
supersymmetric gauge theories. There one introduces an S-duality domain
wall\cite{DGG,HLP} along an $S^3$ inside the $S^4$ which supports a 3D
gauge theory on its worldvolume. AGT relation then implies that the
partition function of the wall theory on $S^3$ should agree with the matrix
element of the corresponding S-duality transformation in the
representation theory of the (extended) conformal symmetry at $b=1$.
In this setting, it has been found \cite{Hama-HL2} that by deforming the
round $S^3$ into a 3D ellipsoid,
\begin{equation}
 \frac{x_1^2+x_2^2}{\ell^2}+\frac{x_3^2+x_4^2}{\tilde\ell^2}~=~1,
\label{ellip}
\end{equation}
with a suitable background $SO(2)_\text{R}$ gauge field to ensure rigid
supersymmetry, one can change the value of the coupling to $b=(\ell/\tilde\ell)^{1/2}$.
For other recent work on supersymmetric deformations of the round $S^3$
with additional background fields, see
\cite{Dolan-SV, Gadde-Y, Imamura, Imamura-Y, Martelli-PS, Martelli-S}.
The above result in three dimensions implies that the correct
deformation of $S^4$ should be a fibration of the ellipsoid
(\ref{ellip}) over a line segment, because the S-duality wall can then
wrap the 3D fiber anywhere in four dimensions in a supersymmetric
manner. 

In this paper we show that SW theories on the 4D ellipsoids,
\begin{equation}
 \frac{x_0^2}{r^2}+\frac{x_1^2+x_2^2}{\ell^2}+\frac{x_3^2+x_4^2}{\tilde\ell^2}~=~1,
\end{equation}
with some additional background fields, reproduce the 2D Liouville or
Toda CFTs with the coupling $b=(\ell/\tilde\ell)^{1/2}$.
As can be easily guessed from our previous result, the additional fields
include an R-symmetry gauge field which takes values on $SU(2)$ Lie
algebra this time. Moreover, it turns out that the relevant off-shell 4D
${\cal N}=2$ supergravity multiplet contains some more auxiliary fields, and
they also have to take nonzero values to make the background supersymmetric.

The organization of this paper is as follows. After a brief summary of
our notations on 4D spinor calculus, in Section \ref{sec:SUSY} we
present the set of Killing spinor equations, and the action and
supersymmetry of general SW theories on arbitrary curved backgrounds
which support Killing spinors. Then in Section \ref{sec:ellipsoid} we
analyze the Killing spinor equation on ellipsoids. It will be shown
that, by assuming that a Killing spinor on round $S^4$ remains after the
deformation of the metric, one can solve the Killing spinor equation
in favor of the background gauge and auxiliary fields and determine
their form up to some arbitrariness. The square of the supersymmetry
yields an isometry of the ellipsoid which fixes two special points,
i.e. the north and south poles. It is shown that the theory looks near
the two poles like the (anti-)topologically twisted theory with Omega
deformation parameter
$(\epsilon_1,\epsilon_2)=(\ell^{-1},\tilde\ell^{-1})$. In Section
\ref{sec:localization} we carry out the explicit path integration using
the SUSY localization principle. Our argument here follows closely that
of Pestun \cite{Pestun}. Finally in Section \ref{sec:conclusion} we
conclude with a few remarks, including a quick check of the AGT relation
in the simplest examples.

\paragraph{Notations.}

Under the 4D rotation group $SO(4)\simeq SU(2)\times SU(2)$, chiral
and anti-chiral spinors transform as doublets of the first and the
second $SU(2)$, respectively. We use the indices $\alpha,\beta,\cdots$
and $\dot\alpha,\dot\beta,\cdots$ for chiral and anti-chiral spinors.
These indices are raised and lowered by the antisymmetric
invariant tensors
$\epsilon^{\alpha\beta},\epsilon^{\dot\alpha\dot\beta},
 \epsilon_{\alpha\beta},\epsilon_{\dot\alpha\dot\beta}$
with nonzero elements
\begin{equation}
 \epsilon^{12}=-\epsilon^{21}=-\epsilon_{12}=\epsilon_{21}=1.
\label{defep}
\end{equation}
Following Wess and Bagger, pairs of undotted indices are suppressed
when contracted in the up-left, down-right order, and similarly for
contracted dotted indices in the down-left, up-right order.

We introduce a set of $2\times 2$ matrices $(\sigma^a)_{\alpha\dot\alpha}$
and $(\bar\sigma^a)^{\dot\alpha\alpha}$ with $a=1,\cdots,4$ satisfying
standard algebras. In terms of Pauli's matrices $\pauli^a$
they are given by
\begin{equation}
\begin{array}{rclrcll}
 \sigma^a&=&-i\pauli^a,\quad&
 \bar\sigma^a&=&i\pauli^a,\quad&(a=1,2,3)\\
 \sigma^4&=&1,&
 \bar\sigma^4&=&1.
\end{array}
\end{equation}
We also use
$\sigma_{ab}=\frac12(\sigma_a\bar\sigma_b-\sigma_b\bar\sigma_a)$ and
$\bar\sigma_{ab}=\frac12(\bar\sigma_a\sigma_b-\bar\sigma_b\sigma_a)$.
Note that $\sigma_{ab}$ is anti self-dual, namely
$\sigma_{ab}=-\frac12\varepsilon_{abcd}\sigma^{cd}$, while
$\bar\sigma_{ab}$ is self-dual.

\section{Seiberg-Witten Theories on Curved Spaces}\label{sec:SUSY}

Manifolds which can support supersymmetric field theories are
characterized by the existence of Killing spinors. In this paper we
consider theories which, when realized on a flat $\mathbb R^4$, have
eight supercharges, i.e. 4D ${\cal N}=2$ supersymmetric theories or
Seiberg-Witten (SW) theories. For these theories, supersymmetry is
characterized by a pair of a chiral and an anti-chiral Killing spinors
$\xi\equiv(\xi_{\alpha A}, \bar\xi^{\dot\alpha}_{\;A})$, both with an
additional $SU(2)_\text{R}$ doublet index $A,B,\cdots$. We use the
tensors $\epsilon^{AB}, \epsilon_{AB}$ with nonzero elements
(\ref{defep}) to raise or lower $SU(2)_\text{R}$ indices. We also
require the Killing spinors to satisfy the reality condition
\begin{equation}
 (\xi_{\alpha A})^\dagger= \xi^{A\alpha}
 =\epsilon^{\alpha\beta}\epsilon^{AB}\xi_{\beta B},\quad
 (\bar\xi_{\dot\alpha A})^\dagger= \bar\xi^{A\dot\alpha}
 =\epsilon^{\dot\alpha\dot\beta}\epsilon^{AB}\xi_{\dot\beta B}.
\label{xireal}
\end{equation}

Our Killing spinor equation consists of two sets of equations. The first
set is called the {\bf main equation}
\begin{eqnarray}
 D_m\xi_A+T^{kl}\sigma_{kl}\sigma_m\bar\xi_A &=&
 -i\sigma_m\bar\xi'_A,
 \nonumber \\
 D_m\bar\xi_A+\bar T^{kl}\bar\sigma_{kl}\bar\sigma_m\xi_A &=&
 -i\bar\sigma_m\xi'_A
 \quad\text{for some }\xi'_A,\bar\xi'_A\,.
\label{ks1}
\end{eqnarray}
Here $T^{kl}, \bar T^{kl}$ are a self-dual and an anti-self-dual real
tensor background fields, and the covariant derivatives contain a
background $SU(2)_\text{R}$ gauge field ${V_m}^A_{~B}$ in addition to
spin connection $\Omega_m^{ab}$.
\begin{eqnarray}
 D_m\xi_A&\equiv& \partial_m\xi_A+\frac14\Omega_m^{ab}\sigma_{ab}\xi_A
 +i\xi_B{V_m}^B_{~A},
 \nonumber \\
 D_m\bar\xi_A&\equiv& \partial_m\bar\xi_A
 +\frac14\Omega_m^{ab}\bar\sigma_{ab}\bar\xi_A +i\bar\xi_B{V_m}^B_{~A}.
\end{eqnarray}
The second set is called the {\bf auxiliary equation}:
\begin{eqnarray}
 \sigma^m\bar\sigma^nD_mD_n\xi_A
 +4D_lT_{mn}\sigma^{mn}\sigma^l\bar\xi_A
 &=& M\xi_A,\nonumber \\
 \bar\sigma^m\sigma^nD_mD_n\bar\xi_A
 +4D_l\bar T_{mn}\bar\sigma^{mn}\bar\sigma^l\xi_A
 &=& M\bar\xi_A,
\label{ks2}
\end{eqnarray}
where $M$ is a scalar background field. We will later show that, if a 4D
manifold with possibly nonzero background fields
$T^{kl},\bar T^{kl}, {V_m}^A_{~B}$ and $M$ admits a Killing spinor
satisfying these equations, one can define SW theories on it with
a rigid supersymmetry.

The above generalized Killing spinor equation was found following the
suggestion of \cite{Festuccia-S} to consider the coupling to off-shell
supergravity. The set of background fields and Killing spinor equations
can be compared to the auxiliary fields in the supergravity multiplet
and BPS equations of \cite{sugra}, but there are some differences due to
the change in spacetime signature. As an example, although SW theories
are known to have $SU(2)\times U(1)$ R-symmetry, we do not consider
background $U(1)_\text{R}$ gauge field because the $U(1)$ phase rotation
is not compatible with the reality condition of SUSY parameter
(\ref{xireal}). Also, this $U(1)_\text{R}$ will be broken explicitly if
the background fields $T^{kl},\bar T^{kl}$ take nonzero values.

\paragraph{Vector multiplets.}

Vector multiplet consists of a gauge field $A_m$, gauginos
$\lambda_{\alpha A}, \bar\lambda_{\dot\alpha A}$, two scalar fields
$\phi,\bar\phi$ and an auxiliary field $D_{AB}=D_{BA}$ all taking values on
the same Lie algebra. Their SUSY transformation rule is given by
\begin{eqnarray}
 \susy A_m &=& i\xi^A\sigma_m\bar\lambda_A-i\bar\xi^A\bar\sigma_m\lambda_A,
 \nonumber \\
 \susy\phi &=& -i\xi^A\lambda_A,
 \nonumber \\
 \susy\bar\phi &=& +i\bar\xi^A\bar\lambda_A,
 \nonumber \\
 \susy\lambda_A &=&
 \tfrac12\sigma^{mn}\xi_A(F_{mn}+8\bar\phi T_{mn})
 +2\sigma^m\bar\xi_AD_m\phi
 +\sigma^mD_m\bar\xi_A\phi+2i\xi_A[\phi,\bar\phi]
 +D_{AB}\xi^B,
 \nonumber \\
 \susy\bar\lambda_A &=&
 \tfrac12\bar\sigma^{mn}\bar\xi_A(F_{mn}+8\phi\bar T_{mn})
 +2\bar\sigma^m\xi_AD_m\bar\phi
 +\bar\sigma^mD_m\xi_A\bar\phi-2i\bar\xi_A[\phi,\bar\phi]+D_{AB}\bar\xi^B,
 \nonumber \\
 \susy D_{AB} &=&
 -i\bar\xi_A\bar\sigma^mD_m\lambda_B
 -i\bar\xi_B\bar\sigma^mD_m\lambda_A
 +i\xi_A\sigma^mD_m\bar\lambda_B
 +i\xi_B\sigma^mD_m\bar\lambda_A
 \nonumber \\ &&
 -2[\phi,\bar\xi_A\bar\lambda_B+\bar\xi_B\bar\lambda_A]
 +2[\bar\phi,\xi_A\lambda_B+\xi_B\lambda_A].
\label{susyvec}
\end{eqnarray}
Here and throughout this paper, we take the Killing spinor $\xi$ to be
Grassmann-even so that $\susy$ is the supercharge which flips the
statistics of the fields. The above transformation rule is compatible
with the reality condition of SUSY parameter (\ref{xireal}) if we assume
\begin{eqnarray}
&&
 (A_m)^\dag=A_m,\quad
 (\lambda_{\alpha A})^\dag=\lambda^{\alpha A},\quad
 (\bar\lambda_{\dot\alpha A})^\dag=\lambda^{\dot\alpha A},
\nonumber \\ &&
 \phi^\dag=\phi,\quad
 (\bar\phi)^\dag=\bar\phi,\quad
 (D_{AB})^\dag=D^{AB}.
\label{vecreal}
\end{eqnarray}
Note that $\phi,\bar\phi$ are two independent real scalar fields.

The supersymmetry algebra closes off-shell, i.e.
$\{\susy_\xi,\susy_\eta\}$ is a sum of bosonic symmetries for arbitrary
pair of Killing spinors $\xi,\eta$. Here we give the formula for the
square $\susy^2$ of the supersymmetry for a Killing spinor $\xi$,
\begin{eqnarray}
 \susy^2A_m &=& iv^nF_{nm}+D_m\Phi,
 \nonumber \\ \,
 \susy^2\phi &=&
 iv^nD_n\phi+i[\Phi,\phi]+(w+2\Theta)\phi,
 \nonumber \\ \,
 \susy^2\bar\phi &=&
 iv^nD_n\bar\phi+i[\Phi,\bar\phi]+(w-2\Theta)\bar\phi,
 \nonumber \\ \,
 \susy^2\lambda_A &=&
 iv^nD_n\lambda_A+i[\Phi,\lambda_A]+(\tfrac32w+\Theta)\lambda_A
 +\tfrac i4\sigma^{kl}\lambda_A D_kv_l+\Theta_{AB}\lambda^B,
 \nonumber \\ \,
 \susy^2\bar\lambda_A &=&
 iv^nD_n\bar\lambda_A+i[\Phi,\bar\lambda_A]+(\tfrac32w-\Theta)\bar\lambda_A
 +\tfrac i4\bar\sigma^{kl}\bar\lambda_A D_kv_l+\Theta_{AB}\bar\lambda^B,
 \nonumber \\ \,
 \susy^2D_{AB} &=&
 iv^nD_n D_{AB}+i[\Phi,D_{AB}]+2w D_{AB}
 +\Theta_{AC} D^C_{~B}+\Theta_{BC} D^C_{~A},
\label{susy2vec}
\end{eqnarray}
where the various transformation parameters are defined as follows,
\begin{eqnarray}
 v^m &=& 2\bar\xi^A\bar\sigma^m\xi_A,
 \nonumber \\
 \Phi &=& -2i\phi\bar\xi^A\bar\xi_A+2i\bar\phi\xi^A\xi_A,
 \nonumber \\
 w &=& -\tfrac i2(\xi^A\sigma^mD_m\bar\xi_A+D_m\xi^A\sigma^m\bar\xi_A),
 \nonumber \\
 \Theta &=& -\tfrac i4(\xi^A\sigma^mD_m\bar\xi_A-D_m\xi^A\sigma^m\bar\xi_A),
 \nonumber \\
 \Theta_{AB} &=&
 -i\xi_{(A}\sigma^mD_m\bar\xi_{B)}+iD_m\xi_{(A}\sigma^m\bar\xi_{B)}.
\label{susyalg1}
\end{eqnarray}
We note that, if $\xi$ satisfies the main Killing spinor equation
(\ref{ks1}) only, the algebra does not close on $D_{AB}$. The failure
term
\begin{eqnarray}
 \Delta_{AB} &=&
-2i\phi(
  \bar\xi_{(A}\bar\sigma^k\sigma^lD_kD_l\bar\xi_{B)}
+4\bar\xi_{(A}\bar\sigma^{mn}\bar\sigma^k\xi_{B)}D_k\bar T_{mn})
\nonumber \\ &&
+2i\bar\phi(
  \xi_{(A}\sigma^k\bar\sigma^lD_kD_l\xi_{B)}
+4\xi_{(A}\sigma^{mn}\sigma^k\bar\xi_{B)}D_kT_{mn}),
\end{eqnarray}
vanishes if $\xi$ satisfies also the auxiliary equation.

The supersymmetric Yang-Mills Lagrangian is given by
\begin{eqnarray}
{\cal L}_\text{YM} &=&
\text{Tr}\Big[
 \frac12F_{mn}F^{mn}
+16F_{mn}(\bar\phi T^{mn}+\phi\bar T^{mn})
+64\bar\phi^2 T_{mn}T^{mn}+64\phi^2\bar T_{mn}\bar T^{mn}
 \nonumber \\ &&\qquad
 -4D_m\bar\phi D^m\phi+2M\bar\phi\phi
 -2i\lambda^A\sigma^mD_m\bar\lambda_A
 -2\lambda^A[\bar\phi,\lambda_A]
 +2\bar\lambda^A[\phi,\bar\lambda_A]
\nonumber \\ && \qquad
 +4[\phi,\bar\phi]^2
 -\frac12D^{AB}D_{AB}
 \Big].
\label{LYM}
\end{eqnarray}
For round $S^4$ of radius $\ell$ with no background $SU(2)_\text{R}$
gauge field or auxiliary tensor fields turned on, this Lagrangian
reduces to the one found by Pestun \cite{Pestun} with
$M=-\frac13R=-\frac4{\ell^2}$. The action is then defined by combining
${\cal L}_\text{YM}$ with the topological term,
\begin{equation}
 S_\text{YM}~=~ \frac1{g_\text{YM}^2}\int \td^4x\sqrt{g}{\cal L}_\text{YM}
 +\frac{i\theta}{8\pi^2}\int\text{Tr}\big(F\wedge F\big).
\end{equation}
Instantons and anti-instantons are topologically non-trivial
configurations of gauge field satisfying $\ast F=-F$ or $\ast F=F$,
and are characterized by the instanton number $n\in\mathbb Z$.
The classical action on instanton or anti-instanton backgrounds
takes values
\begin{equation}
 \begin{array}{rcl}
 \text{instanton } (n>0) &:& -S_\text{YM}=2\pi in\tau\,,\\
 \text{anti-instanton } (n<0) &:& -S_\text{YM}=2\pi in\bar\tau\,,
 \end{array}
\quad
 \tau\equiv\frac\theta{2\pi}+\frac{4\pi i}{g_\text{YM}^2}\,.
\end{equation}

The Lagrangian (\ref{LYM}) is not positive definite and path integral
becomes ill-defined if the fields take values according to the reality
condition (\ref{vecreal}). The actual path integral should therefore be
defined with the modified contours along which
\begin{equation}
 \phi^\dagger = -\bar\phi,\quad
 (D_{AB})^\dagger = -D^{AB}.
\label{realityvec}
\end{equation}

For $U(1)$ gauge group, there is also the Fayet-Iliopoulos type
invariant. Let $w^{AB}=w^{BA}$ be a $SU(2)_\text{R}$ triplet background
field satisfying
\begin{eqnarray}
 w^{AB}\xi_B&=& \frac12\sigma^nD_n\bar\xi^A+2T_{kl}\sigma^{kl}\xi^A,
 \nonumber \\
 w^{AB}\bar\xi_B&=& \frac12\bar\sigma^nD_n\xi^A
 +2\bar T_{kl}\bar\sigma^{kl}\bar\xi^A.
\label{FIcoeff}
\end{eqnarray}
Then one can construct the following invariant from a $U(1)$ vector multiplet,
\begin{equation}
 {\cal L}_\text{FI}~\equiv~
 w^{AB}D_{AB}-M(\phi+\bar\phi)-64\phi T^{kl}T_{kl}
 -64\bar\phi\bar T^{kl}\bar T_{kl}-8F^{kl}(T_{kl}+\bar T_{kl})\,.
\label{LFI}
\end{equation}

\paragraph{Hypermultiplets.}

The system of $r$ hypermultiplets consists of scalars $q_{AI}$ and
fermions $\psi_{\alpha I}, \bar\psi^{\dot\alpha}_I$ satisfying the
reality conditions
\begin{eqnarray}
&& (q_{IA})^\dagger\,=q^{AI}\;=\Omega^{IJ}\epsilon^{AB}q_{JB}\,,
\nonumber \\
&& (\psi_{\alpha I})^\dagger=\psi^{\alpha I}
 =\epsilon^{\alpha\beta}\Omega^{IJ}\psi_{\beta J}\,,
 \nonumber\\
&& (\bar\psi_{\dot\alpha I})^\dagger=\bar\psi^{\dot\alpha I}
 =\epsilon^{\dot\alpha\dot\beta}\Omega^{IJ}\bar\psi_{\dot\beta J}\,.
\label{qpsireal}
\end{eqnarray}
Here $I,J=1,\cdots,2r$ are $Sp(r)$ indices and $\Omega^{IJ}$ is the real
antisymmetric $Sp(r)$-invariant tensor satisfying
\begin{equation}
 (\Omega^{IJ})^\ast=-\Omega_{IJ},\quad
 \Omega^{IJ}\Omega_{JK}=\delta^I_K.
\end{equation}
Pairs of $Sp(r)$ indices contracted in the order of top-left,
bottom-right will be often suppressed. For example,
$q^Aq_A\equiv q^{AI}q_{IA}$.
These matter fields can couple to vector multiplets through an embedding of
the gauge group into $Sp(r)$. Namely, when vector multiplet fields such
as $A_m$ are multiplied on hypermultiplet fields, they are thought of as
$2r\times 2r$ matrices with elements $(A_m)_I^{~J}$. The covariant
derivatives of matters therefore take the form
\begin{eqnarray}
 D_mq_{IA}&\equiv&\partial_mq_{IA}-i(A_m)_I^{~J}q_{JA}
 +iq_{IB}(V_m)^B_{~A},
 \nonumber \\
 D_m\psi_{\alpha I}&\equiv& \partial_m\psi_{\alpha I}
 -i(A_m)_I^{~J}\psi_{\alpha J}
 +\frac14\Omega_m^{ab}(\sigma_{ab})_\alpha^{~\beta}\psi_{\beta I}
 ,\quad\text{etc.}
\end{eqnarray}

It is straightforward to find on-shell SUSY transformation rule,
\begin{eqnarray}
 \susy^\text{os}q_A &=& -i\xi_A\psi+i\bar\xi_A\bar\psi,
 \nonumber \\
 \susy^\text{os}\psi &=&
 2\sigma^m\bar\xi_AD_mq^A+\sigma^mD_m\bar\xi_Aq^A
 -4i\xi_A\bar\phi q^A, \nonumber \\
 \susy^\text{os}\bar\psi &=&
 2\bar\sigma^m\xi_AD_mq^A+\bar\sigma^mD_m\xi_Aq^A-4i\bar\xi_A\phi q^A,
\end{eqnarray}
and the gauge covariant kinetic Lagrangian
\begin{eqnarray}
{\cal L}_\text{mat}^\text{os} &=&
 \frac12D_mq^AD^mq_A
 -q^A\{\phi,\bar\phi\}q_A
 +\frac i2q^AD_{AB}q^B
 +\frac18\left(R+M\right)q^Aq_A
 \nonumber \\ &&
 -\frac i2\bar\psi\bar\sigma^mD_m\psi
 -\frac12\psi\phi\psi +\frac12\bar\psi\bar\phi\bar\psi
 +\frac i2\psi\sigma^{kl}T_{kl}\psi
 -\frac i2\bar\psi\bar\sigma^{kl}\bar T_{kl}\bar\psi
 \nonumber \\ &&
 -q^A\lambda_A\psi+\bar\psi\bar\lambda_Aq^A.
\end{eqnarray}
It is known that one cannot make the full ${\cal N}=2$ SUSY
transformation law closed off-shell with finitely many auxiliary fields.
For the application of localization principle, however, one focuses on the
supersymmetry $\susy$ corresponding to a specific choice of Killing spinor
$\xi$. It is then sufficient that $\susy^2$ for that specific $\xi$
is a linear sum of bosonic symmetries on all fields off-shell.

We introduce the auxiliary scalars $F_{AI}$ satisfying the reality
condition
\begin{equation}
 (F_{IA})^\dagger=F^{AI}=\Omega^{IJ}\epsilon^{AB}F_{JB}\,,
\label{Freal}
\end{equation}
and put the full Lagrangian as follows,
\begin{equation}
 {\cal L}_\text{mat}~=~ {\cal L}_\text{mat}^\text{os}-\frac12F^AF_A.
\label{Lmatful}
\end{equation}
The supersymmetry transformation laws of fields are extended as follows,
\begin{eqnarray}
 \susy q_A &=& -i\xi_A\psi+i\bar\xi_A\bar\psi,
 \nonumber \\
 \susy\psi &=&
 2\sigma^m\bar\xi_AD_mq^A+\sigma^mD_m\bar\xi_Aq^A
 -4i\xi_A\bar\phi q^A +2\check\xi_AF^A, \nonumber \\
 \susy\bar\psi &=&
 2\bar\sigma^m\xi_AD_mq^A+\bar\sigma^mD_m\xi_Aq^A-4i\bar\xi_A\phi q^A
 +2\bar{\check\xi}_AF^A,
 \nonumber \\
 \susy F_A &=&
 +i\check\xi_A\sigma^mD_m\bar\psi
 -2\check\xi_A\phi\psi
 -2\check\xi_A\lambda_Bq^B
 +2i\check\xi_A(\sigma^{kl}T_{kl})\psi
 \nonumber \\ &&
 -i\bar{\check\xi}_A\bar\sigma^mD_m\psi
 +2\bar{\check\xi}_A\bar\phi\bar\psi
 +2\bar{\check\xi}_A\bar\lambda_Bq^B
 -2i\bar{\check\xi}_A(\bar\sigma^{kl}\bar T_{kl})\bar\psi.
\label{susyhyp}
\end{eqnarray}
Here the new transformation parameters $\check\xi,\bar{\check\xi}$
are required to satisfy
\begin{eqnarray}
 \xi_A\check\xi_B-\bar\xi_A\bar{\check\xi}_B &=& 0,
 \nonumber \\
 \xi^A\xi_A+\bar{\check\xi}^A\bar{\check\xi}_A &=& 0,
 \nonumber \\
 \bar\xi^A\bar\xi_A+\check\xi^A\check\xi_A &=& 0,
 \nonumber \\
 \xi^A\sigma^m\bar\xi_A+\check\xi^A\sigma^m\bar{\check\xi}_A &=&0.
\label{checkxi}
\end{eqnarray}
Similar off-shell transformation rule which makes use of constrained
transformation parameters like $\check\xi,\bar{\check\xi}$ here has been
written down for 4D ${\cal N}=4$ gauge theories on $S^4$ in \cite{Pestun}, and
for 5D SUSY theories on $S^5$ in \cite{HST}. One can then show that
$\susy$ squares into a linear sum of bosonic symmetries off-shell,
\begin{eqnarray}
 \susy^2 q_A &=&
  iv^mD_mq_A+i\Phi q_A+wq_A
  +\Theta_{AB}q^B ,
 \nonumber \\
 \susy^2\psi &=&
 iv^mD_m\psi+i\Phi\psi+\tfrac32w\psi-\Theta\psi
 +\tfrac i4\sigma^{kl}\psi D_{k}v_{l},
 \nonumber \\
 \susy^2\bar\psi &=&
 iv^mD_m\bar\psi+i\Phi\bar\psi+\tfrac32w\bar\psi+\Theta\psi
 +\tfrac i4\bar\sigma^{kl}\bar\psi D_{k}v_{l},
 \nonumber \\
 \susy^2 F_A &=&
 iv^mD_m F_A+i\Phi F_A+2wF_A
 +\check\Theta_{AB}F^B.
\label{susy2hyp}
\end{eqnarray}
Here the parameters $v^m,\Phi,w,\Theta,\Theta_{AB}$ are as in
(\ref{susyalg1}) and
\begin{equation}
 \check\Theta_{AB} ~=~
 2i\check\xi_{(A}\sigma^mD_m\bar{\check\xi}_{B)}
 -2iD_m\check\xi_{(A}\sigma^m\bar{\check\xi}_{B)}
 +4i\check\xi_{(A}\sigma^{kl}T_{kl}\check\xi_{B)}
 -4i\bar{\check\xi}_{(A}\bar\sigma^{kl}\bar T_{kl}\bar{\check\xi}_{B)}.
\label{susyalg2}
\end{equation}
Note that $\check\xi_A,\bar{\check\xi}_A$ and $F_A$ transform as
doublets under a local symmetry which we call
$SU(2)_{\check{\text{R}}}$, reflecting the fact that the choice of
$\check\xi_A,\bar{\check\xi}_A$ satisfying (\ref{checkxi}) is not unique.
This also means that the covariant derivative of $F_A$ contains the
background $SU(2)_{\check{\text{R}}}$ gauge field $\check V_{m~~A}^{~~B}$.
\begin{equation}
 D_mF_{IA}~\equiv~ \partial_mF_{IA} -i(A_m)_{I}^{~J}F_{JA}
 +iF_{IB}\check V_{m~A}^{~\,B}.
\end{equation}
For notational simplicity, we use for their doublet indices the same
letters $A,B,\cdots$ as for the $SU(2)_\text{R}$ indices.

The off-shell transformation rule (\ref{susyhyp}) is compatible with the
reality condition of the fields (\ref{qpsireal}) and (\ref{Freal}).
However, if we define the theory of hypermultiplets by the kinetic
Lagrangian ${\cal L}_\text{mat}$, we have to take the actual path
integration contour in such a way that its bosonic part is positive
definite. Therefore, we choose the integration contour for $F_{AI}$
differently from its real locus, so that
\begin{equation}
 (F_{IA})^\dagger=-F^{AI}
\end{equation}
along the contour.

An important fact which will be used later is that the matter kinetic
Lagrangian is supersymmetry exact. Assuming
$\xi^A\xi_A-\bar\xi^A\bar\xi_A=1$ which will be verified in the next
section, one can show that
\begin{eqnarray}
 {\cal L}_\text{mat} &=& \susy {\cal V}_\text{mat},
 \nonumber \\
 2{\cal V}_\text{mat} &=&
  \psi\check\xi^AF_A
 -\bar\psi\bar{\check\xi}^AF_A
 +\psi\sigma^mD_m(\bar\xi_Aq^A)
 -\bar\psi\bar\sigma^mD_m(\xi_Aq^A)
 \nonumber \\ &&
 +2(i\psi\phi+\psi\sigma^{kl}T_{kl}+iq^B\lambda_B)\xi_Aq^A
 -2(i\bar\psi\bar\phi+\bar\psi\bar\sigma^{kl}\bar T_{kl}
  +iq^B\bar\lambda_B)\bar\xi_Aq^A.
\end{eqnarray}

\section{Supersymmetry on 4D Ellipsoids}\label{sec:ellipsoid}

It has been known that round spheres in various dimensions admit Killing
spinors satisfying
\begin{equation}
  D_m\zeta ~=~ \Gamma_m\zeta'~~\text{for some }\zeta'.
\label{ks-gen}
\end{equation}
In \cite{Hama-HL2} it was shown that the 3D ellipsoids with
$U(1)\times U(1)$ isometry admit a pair of charged Killing spinors
coupled to a suitably chosen background $U(1)_\text{R}$ gauge field. The
ellipsoid is defined by an embedding equation in flat $\mathbb R^4$ with
Cartesian coordinates $x_1,\cdots,x_4$,
\begin{equation}
 \frac{x_1^2+x_2^2}{\ell^2}+\frac{x_3^2+x_4^2}{\tilde\ell^2}~=~1.
\label{3Dellipsoids}
\end{equation}
The goal of this section is to show that similar ellipsoids in four dimensions,
\begin{equation}
 \frac{x_0^2}{r^2}+\frac{x_1^2+x_2^2}{\ell^2}
 +\frac{x_3^2+x_4^2}{\tilde\ell^2}=1,
\label{ellipsoids}
\end{equation}
admit a Killing spinor satisfying (\ref{ks1}) and (\ref{ks2}) if the
background fields ${V_m}^A_{~B}, T_{kl},\bar T_{kl}, M$ are chosen
appropriately. We will restrict to those backgrounds with
$U(1)\times U(1)$ isometry, and anticipate that the square of the
supersymmetry yield a linear combination of the two $U(1)$ isometries
which fix the north and south poles of the ellipsoid. Our ellipsoid
(\ref{ellipsoids}) is thus parametrized by three axis-length parameters.

Introducing a polar coordinate system,
\begin{equation}
\begin{array}{rcl}
 x_0 &=& r\cos\rho, \\
 x_1 &=& \ell\sin\rho\,\cos\theta\,\cos\varphi, \\
 x_2 &=& \ell\sin\rho\,\cos\theta\,\sin\varphi, \\
 x_3 &=& \tilde\ell\sin\rho\,\sin\theta\,\cos\chi, \\
 x_4 &=& \tilde\ell\sin\rho\,\sin\theta\,\sin\chi,
\end{array}
\end{equation}
the vielbein one-forms $E^a=E^a_m\td x^m$ can be chosen as
\begin{equation}
 E^1=\sin\rho e^1,\quad
 E^2=\sin\rho e^2,\quad
 E^3=\sin\rho e^3+h \td \rho,\quad
 E^4=g \td \rho,
\end{equation}
where
\begin{eqnarray}
 f&\equiv& \sqrt{\ell^2\sin^2\theta+\tilde\ell^2\cos^2\theta},
 \nonumber \\
 g&\equiv& \sqrt{r^2\sin^2\rho+\ell^2\tilde\ell^2f^{-2}\cos^2\rho},
 \nonumber \\
 h&\equiv& \frac{\tilde\ell^2-\ell^2}f\cos\rho\sin\theta\cos\theta,
\label{fgh}
\end{eqnarray}
and $e^a$ are vielbein of the 3D ellipsoid (\ref{3Dellipsoids}) in
polar coordinates $(\varphi,\chi,\theta)$,
\begin{equation}
 e^1=\ell\cos\theta \td \varphi,\quad
 e^2=\tilde\ell\sin\theta \td \chi,\quad
 e^3=f \td \theta.
\end{equation}
The spin connection $\Omega^{ab}=\Omega^{ab}_m\td x^m$ has the following components,
\begin{eqnarray}
&&
 \Omega^{12}=0,\quad
 \Omega^{13}=-\frac\ell f\sin\theta \td \varphi,\quad
 \Omega^{23}=\frac{\tilde\ell}f\cos\theta \td \chi,\nonumber \\
&&
 \Omega^{14}=\frac{\tilde\ell^2\cos\rho}{gf^2}e^1,\quad
 \Omega^{24}=\frac{\ell^2\cos\rho}{gf^2}e^2,\quad
 \Omega^{34}=\frac{\ell^2\tilde\ell^2\cos\rho}{gf^4}e^3.
\end{eqnarray}
Note that $\Omega^{12},\Omega^{13},\Omega^{23}$ are the spin connection
of the 3D ellipsoid with vielbein $e^a$.

\paragraph{Killing spinors on round $\bf S^4$.}

Killing spinor equation has solutions on the round $S^4$ of radius
$\ell$ with no background gauge or tensor auxiliary fields turned on. The
main equation (\ref{ks1}) consists of eight equations, and we divide
them into two groups. The first six equations are given by
\begin{eqnarray}
 \Big(\partial_m+\frac14\Omega_m^{ab}\pauli^{ab}
 -\frac{i\cos\rho}{2\ell}e^a_m\pauli^a\Big)\xi_A
 &=& -\sin\rho e^a_m\pauli^a\bar\xi'_A,
 \nonumber \\
 \Big(\partial_m+\frac14\Omega_m^{ab}\pauli^{ab}
 +\frac{i\cos\rho}{2\ell}e^a_m\pauli^a\Big)\bar\xi_A
 &=& +\sin\rho e^a_m\pauli^a\xi'_A,
\label{KS01}
\end{eqnarray}
where $a,b=1,2,3$ and the index $m$ runs over
$\varphi,\chi,\theta$. Here $\pauli^a$ are Pauli's matrices as before
and we used $\pauli^{ab}\equiv\frac12(\pauli^a\pauli^b-\pauli^b\pauli^a)$.
The last two equations read
\begin{eqnarray}
 \partial_\rho\xi_A &=& -i\ell\bar\xi'_A, \nonumber \\
 \partial_\rho\bar\xi_A &=& -i\ell\xi'_A.
\label{KS02}
\end{eqnarray}
The equations (\ref{KS01}) are solved by Killing spinors
$\kappa_{st}~(s,t=\pm1)$ on round $S^3$ of radius $\ell$ with
coordinates $\theta,\varphi,\chi$, which satisfy
\begin{equation}
 \Big(\partial_m+\frac14\Omega_m^{ab}\pauli^{ab}\Big)\kappa_{st}=
 -\frac{ist}{2\ell}e^a_m\pauli^a\kappa_{st},\quad
 \kappa_{st}\equiv\frac12\left(\begin{array}{r}
   e^{\frac i2(s\chi+t\varphi-st\theta)} \\
 -se^{\frac i2(s\chi+t\varphi+st\theta)}
 \end{array}\right)
\end{equation}
for $m=(\varphi,\chi,\theta)$ and $a,b=1,2,3$. One can form a Killing
vector on the $S^3$ as a bilinear of $\kappa_{st}$,
\begin{equation}
 \kappa_{st}^\dagger \pauli_a \kappa_{st}\cdot e^{am}\partial_m
 ~=~ -\frac1{2\ell}(s\partial_\varphi+t\partial_\chi).
\end{equation}
Recalling that $\partial_\varphi$ and $\partial_\chi$ are rotations in
the $(x_1,x_2)$-plane and $(x_3,x_4)$-plane, we restrict to those with
$s=t$ so that our choice of Killing spinor corresponds to Omega
deformations with $\epsilon_1=\epsilon_2$ at the north pole. Assuming
$\xi_A$ and $\bar\xi_A$ are all proportional to $\kappa_{_{++}}$ or
$\kappa_{_{--}}$, the remaining equations (\ref{KS02}) become
\begin{eqnarray}
 -i\ell\bar\xi'_A &=& \partial_\rho\xi_A
  ~=~ \frac{\cos\rho+1}{2\sin\rho}\xi_A,
 \nonumber \\
 -i\ell\xi'_A &=& \partial_\rho\bar\xi_A
  ~=~ \frac{\cos\rho-1}{2\sin\rho}\bar\xi_A.
\end{eqnarray}

In the following we take a particular solution of the main equation
(\ref{ks1}) which also satisfies the reality condition (\ref{xireal}).
We also require
\begin{equation}
 \xi^A\xi'_A~=~ \bar\xi^A\bar\xi'_A~=~0,
\label{wr0}
\end{equation}
so that the square of the corresponding supersymmetry transformation
does not give rise to dilation or $U(1)_\text{R}$ transformation,
namely $w=\Theta=0$ in (\ref{susyalg1}). It is unique up to the symmetries
of the theory.
\begin{eqnarray}
 \xi_A
 ~=~\left(\xi_1,\xi_2\right)
 &=&\sin\frac\rho2
  \left(\kappa_{_{++}},\kappa_{_{--}}\right),
 \nonumber \\
 \bar\xi_A
 ~=~\left(\bar\xi_1,\bar\xi_2\right)
 &=&\cos\frac\rho2
  \left(i\kappa_{_{++}},-i\kappa_{_{--}}\right).
\label{ourks}
\end{eqnarray}
The Killing vector which appears in the square of the supersymmetry
transformation is
\begin{equation}
 v^m\partial_m ~=~
 2\bar\xi^A\bar\sigma^m\xi_A\,\partial_m ~=~ \frac 1\ell
 (\partial_\varphi+\partial_\chi).
\label{kv1}
\end{equation}
This solution also satisfies the auxiliary equation (\ref{ks2})
with the choice $M=-\frac13R=-4\ell^{-2}$.

\paragraph{A Killing spinor on ellipsoids.}

Next we study the Killing spinor equation on ellipsoids
(\ref{ellipsoids}). Our strategy is to assume that, for a suitable
choice of the background gauge and auxiliary fields, the Killing spinor
(\ref{ourks}) on round $S^4$ remains a Killing spinor also on
ellipsoids. Then we will see that the Killing spinor equation can be
turned into a set of linear algebraic equations on the background
fields which have nontrivial solutions. A similar approach worked in the
case of 3D ellipsoids \cite{Hama-HL2}. Note that under this assumption
the Killing vector on the ellipsoid becomes
\begin{equation}
 2\bar\xi^A\bar\sigma^m\xi_A\,\partial_m ~=~
  \frac 1\ell\partial_\varphi
 +\frac 1{\tilde\ell}\partial_\chi,
\label{kv2}
\end{equation}
which can be interpreted as the Omega deformation with
$\epsilon_1=\ell^{-1}, \epsilon_2=\tilde\ell^{-1}$ near the north and
south poles. This point will be explained in more detail later.

In solving the Killing spinor equation to determine the background
fields, a useful fact is that the 3D spinors $\kappa_{st}$ on $S^3$ remain
Killing spinors after the deformation to 3D ellipsoids if a suitable
background $U(1)$ gauge field is turned on at the same time. More
explicitly, one has
\begin{eqnarray}
 \Big(\partial_m+\frac14\Omega^{ab}_m\pauli^{ab}\mp iV_m^{[3]}\Big)
 \kappa_{_{\pm\pm}}~=~ -\frac{i}{2f}e^a_m\pauli^a\kappa_{_{\pm\pm}},
 \nonumber \\
 V^{[3]}~\equiv~
 \frac 12\Big(1-\frac\ell f\Big)d\varphi
+\frac 12\Big(1-\frac{\tilde\ell}f\Big)d\chi,
\end{eqnarray}
where $m=\varphi,\chi,\theta$ and $a,b=1,2,3$.
Another useful fact is that the following $2\times 2$ matrix,
\begin{equation}
 \pauli^1_\theta~\equiv \pauli^1\cos\theta+\pauli^2\sin\theta,
\end{equation}
satisfies $\pauli^1_\theta\kappa_{_{\pm\pm}}=\mp\kappa_{_{\pm\pm}}$
and therefore $\pauli^1_\theta\xi_A= -\xi_B(\pauli^3)^B_{~A}$. At this point we
find it convenient to regard $\xi_A$ and $\bar\xi_A$ as $2\times 2$
matrices, on which $2\times 2$ matrices with spinor indices act from
the left and those with $SU(2)_{\rm R}$ indices act from the right.
The latter equation can then be rewritten in the matrix form,
\begin{equation}
 \pauli^1_\theta\bs\xi=-\bs\xi\pauli^3.
\end{equation}
Hereafter all the boldface letters can be regarded as $2\times2$ matrix
quantities. By using the above equation in combination with
\begin{equation}
 \pauli^3\bs\xi=\bs\xi
 \big\{ \cos(\chi+\varphi)\pauli^1+\sin(\chi+\varphi)\pauli^2\big\},
\end{equation}
any $SU(2)$ action from the right of $\bs\xi$ can be translated into an
$SU(2)$ action from the left, and vice versa. Note also that
\begin{equation}
 \pauli^1_\theta\bs\xi=i\tan\frac\rho2\bar{\bs\xi}.
\end{equation}

Let us now turn to the analysis of Killing spinor equation. We introduce the
notations
\begin{equation}
 {\bf V} +V^{[3]}\pauli^3\equiv\tilde{\bf V}= E^a\tilde{\bf V}_a,\quad
 i{\bf T} ~\equiv~ \sigma_{kl}T^{kl},\quad
 i\bar{\bf T} ~\equiv~ \bar\sigma_{kl}\bar T^{kl}.
\end{equation}
We also require that (\ref{wr0}) is still satisfied on ellipsoids, and
introduce a pair of anti-symmetric tensors $S_{kl},\bar S_{kl}$ and
matrices ${\bf S},\bar{\bf S}$ by the formula
\begin{equation}
 \bs\xi'= {\bf S}\bs\xi= -i\sigma_{kl}S^{kl}\bs\xi,\quad
 \bar{\bs\xi}'= \bar{\bf S}\bar{\bs\xi}=
 -i\bar\sigma_{kl}\bar S^{kl}\bar{\bs\xi}.
\end{equation}
Inserting these together with (\ref{ourks}) into the main equation
(\ref{ks1}), we obtain
\begin{eqnarray}
  \bs\xi\tilde{\bf V}_4
 +{\bf T}\bar{\bs\xi}
 +\bar{\bf S}\bar{\bs\xi}
 &=&
 i\frac{\cos\rho+1}{2g\sin\rho}\bs\xi
 -\frac{h}{2fg\sin\rho}\pauli^3\bs\xi
 -\frac{h\Omega^{34}_3}{2g}\pauli^3\bs\xi,
 \nonumber \\
  \bar{\bs\xi}\tilde{\bf V}_4
 +\bar{\bf T}\bs\xi
 +{\bf S}\bs\xi
 &=&
 i\frac{\cos\rho-1}{2g\sin\rho}\bar{\bs\xi}
 -\frac{h}{2fg\sin\rho}\pauli^3\bar{\bs\xi}
 +\frac{h\Omega^{34}_3}{2g}\pauli^3\bar{\bs\xi},
\label{kseq2-1}
\end{eqnarray}
and
\begin{eqnarray}
 \bs\xi\tilde{\bf V}_a
 -i{\bf T}\pauli^a\bar{\bs\xi}
 -i\pauli^a\bar{\bf S}\bar{\bs\xi}
 &=&
  \frac1{2f\sin\rho}\pauli^a\bs\xi
 +\frac12\Omega_a^{b4}\pauli^b\bs\xi,
 \nonumber \\
 \bar{\bs\xi}\tilde{\bf V}_a
 +i\bar{\bf T}\pauli^a\bs\xi
 +i\pauli^a{\bf S}\bs\xi
 &=&
  \frac1{2f\sin\rho}\pauli^a\bar{\bs\xi}
 -\frac12\Omega_a^{b4}\pauli^b\bar{\bs\xi},
\label{kseq2-2}
\end{eqnarray}
where $a,b=1,2,3$ and the nonzero components of $\Omega_a^{b4}$ are
\begin{equation}
 \Omega_1^{14}=\frac{\tilde\ell^2\cos\rho}{gf^2\sin\rho},\quad
 \Omega_2^{24}=\frac{\ell^2\cos\rho}{gf^2\sin\rho},\quad
 \Omega_3^{34}=\frac{\ell^2\tilde\ell^2\cos\rho}{gf^4\sin\rho}.
\end{equation}

The equations (\ref{kseq2-1}) and (\ref{kseq2-2}) can be regarded as a
system of inhomogeneous linear algebraic equations for the unknowns
${\bf\tilde V}, {\bf T},{\bf\bar T},{\bf S}$ and $\bar{\bf S}$. We found
that these equations have nontrivial solutions, and moreover the
solution is not unique. A special solution for which
${\bf T},\bar{\bf T}$ take particularly simple form is
\begin{eqnarray}
 {\bf T} &=& \frac14\Big(\frac1f-\frac1g\Big)\pauli^1_\theta
 +\frac h{4fg}\pauli^2_\theta, \nonumber \\
 \bar{\bf T} &=& \frac14\Big(\frac1f-\frac1g\Big)\pauli^1_\theta
 -\frac h{4fg}\pauli^2_\theta, \nonumber \\
 {\bf S} &=& -\frac14\Big(\frac1f+\frac1g\Big)\pauli^1_\theta
 -\frac h{4fg}\pauli^2_\theta, \nonumber \\
 \bar{\bf S} &=& -\frac14\Big(\frac1f+\frac1g\Big)\pauli^1_\theta
 +\frac h{4fg}\pauli^2_\theta, \nonumber \\
 \bs\xi\tilde{\bf V}_1 &=&
  \bigg\{\frac{\cos\theta}{2\sin\rho}\Big(\frac1f-\frac1g\Big)
        -\frac{\sin\theta\cos\rho}{2\sin\rho}\frac h{fg}\bigg\}\pauli^1_\theta\bs\xi
 +\frac{\sin\theta\cos\rho}{2f\sin\rho}\Big(1-\frac{\tilde\ell^2}{gf}\Big)
 \pauli^2_\theta\bs\xi,
 \nonumber \\
 \bs\xi\tilde{\bf V}_2 &=&
  \bigg\{\frac{\sin\theta}{2\sin\rho}\Big(\frac1f-\frac1g\Big)
        +\frac{\cos\theta\cos\rho}{2\sin\rho}\frac h{fg}\bigg\}\pauli^1_\theta\bs\xi
 -\frac{\cos\theta\cos\rho}{2f\sin\rho}\Big(1-\frac{\ell^2}{gf}\Big)
 \pauli^2_\theta\bs\xi,
 \nonumber \\
 \bs\xi\tilde{\bf V}_3 &=&
 -\frac{\cos\rho}{2f\sin\rho}
  \Big(1-\frac{\ell^2\tilde\ell^2}{gf^3}\Big)\pauli^3\bs\xi,
 \nonumber \\
 \bs\xi\tilde{\bf V}_4 &=&
  \frac{h\cos\rho}{2fg\sin\rho}
  \Big(1-\frac{\ell^2\tilde\ell^2}{gf^3}\Big)\pauli^3\bs\xi.
\label{solsp}
\end{eqnarray}
where we introduced $\pauli^2_\theta\equiv i\pauli^1_\theta\pauli^3$.
This special solution can be shifted by solutions of the homogeneous
equation, namely the equations (\ref{kseq2-1}) and (\ref{kseq2-2})
with the r.h.s. set to zero. They are parametrized by three arbitrary
functions $c_1,c_2,c_3$ as follows.
\begin{eqnarray}
 \Delta{\bf T} &=& \tan\tfrac\rho2
 \left(+c_1\pauli^1_\theta + c_2 \pauli^2_\theta + c_3\pauli^3\right),
 \nonumber \\
 \Delta\bar{\bf T} &=& \cot\tfrac\rho2
 \left(-c_1\pauli^1_\theta + c_2 \pauli^2_\theta + c_3\pauli^3\right),
 \nonumber \\
 \Delta{\bf S} &=& \cot\tfrac\rho2
 \left(+c_1\pauli^1_\theta + c_2 \pauli^2_\theta + c_3\pauli^3\right),
 \nonumber \\
 \Delta\bar{\bf S} &=& \tan\tfrac\rho2
 \left(-c_1\pauli^1_\theta + c_2 \pauli^2_\theta + c_3\pauli^3\right),
 \nonumber \\
 \bs\xi\cdot\Delta\tilde{\bf V}_1 &=&
 -2\sin\theta
 \left( c_2\pauli^1_\theta\bs\xi - c_1\pauli^2_\theta\bs\xi \right),
 \nonumber\\
 \bs\xi\cdot\Delta\tilde{\bf V}_2 &=&
 +2\cos\theta
 \left( c_2\pauli^1_\theta\bs\xi - c_1\pauli^2_\theta\bs\xi \right),
 \nonumber\\
 \bs\xi\cdot\Delta\tilde{\bf V}_3 &=&
  -2c_1\pauli^3\bs\xi+2c_3\pauli^1\bs\xi,
 \nonumber \\
 \bs\xi\cdot\Delta\tilde{\bf V}_4 &=&
  +2c_2\pauli^3\bs\xi-2c_3\pauli^2\bs\xi.
\end{eqnarray}

In $2\times2$ matrix notations, the auxiliary equation (\ref{ks2}) becomes
\begin{eqnarray}
 -4\cot\frac\rho2\Big(\sigma^mD_m\bar{\bf S}-D_m{\bf T}\sigma^m\Big)
 \pauli^1_\theta
 -4\sigma^m\bar{\bf S}\bar{\bf T}\bar\sigma_m
 \nonumber \\ ~=~
 4\tan\frac\rho2\Big(\bar\sigma^mD_m{\bf S}-D_m\bar{\bf T}\bar\sigma^m\Big)
 \pauli^1_\theta
 -4\bar\sigma^m{\bf S}{\bf T}\sigma_m
 &=& M\cdot{\bf 1}.
\end{eqnarray}
This is satisfied by the above special solution (\ref{solsp}) with
\begin{equation}
 M ~=~ \frac1{f^2}-\frac1{g^2}+\frac{h^2}{f^2g^2}-\frac4{fg}.
\end{equation}
We found that the auxiliary equation is still satisfied even after
nonzero $c_1,c_2,c_3$ are turned on, as long as they are functions of
$\theta$ and $\rho$ only. The shift of $M$ is then given by
\begin{eqnarray}
 \Delta M &=&
 8\Big( \frac1g\partial_\rho -\frac{h}{gf\sin\rho}\partial_\theta
   +\frac{\ell^2\tilde\ell^2\cos\rho}{gf^4\sin\rho}
   +\frac{\cos\rho(\ell^2+\tilde\ell^2-f^2)}
         {gf^2\sin\rho}
   -\frac{\cos\rho}{f\sin\rho}\Big)c_1
 \nonumber \\ &&
 +8\Big( \frac1{f\sin\rho}\partial_\theta
   + \frac{h\ell^2\tilde\ell^2\cos\rho}{g^2f^4\sin\rho}
   +\frac{2\cot2\theta}{f\sin\rho}
   -\frac{h\cos\rho}{fg\sin\rho} \Big)c_2
 -16(c_1^2+c_2^2+c_3^2).
\end{eqnarray}

We thus determined the form of all the additional background fields in
order for SW theories on the ellipsoid (\ref{ellipsoids}) to admit a
rigid supersymmetry. In the rest of this section we check two more
properties of our background. The first is that the square of the
supersymmetry is a sum of bosonic transformations which indeed leave all
the background fields invariant. The second is that our background is
regular and approaches Omega background near the two poles.

\paragraph{Square of SUSY.}

The supersymmetry transformation $\susy$ acting on fields of SW theory
squares into a sum of bosonic symmetries according to (\ref{susy2vec})
and (\ref{susy2hyp}). It can also be expressed as
\begin{eqnarray}
 \susy^2 &=& i{\cal L}_v + \text{Gauge}(\hat\Phi)
 +\text{Lorentz}(L_{ab})
 \nonumber \\ &&
 +\text{Scale}(w)
 +\text{R}_{U(1)}(\Theta)
 +\text{R}_{SU(2)}(\hat\Theta_{AB})
 +\check{\text{R}}_{SU(2)}(\hat{\check\Theta}_{AB}),
\end{eqnarray}
where
\begin{eqnarray}
 \hat\Phi &\equiv& \Phi-iv^nA_n, \nonumber \\
 L_{ab} &\equiv& D_{[a}v_{b]}+v^n\Omega_{nab}, \nonumber \\
 \hat\Theta_{AB} &\equiv& \Theta_{AB}+v^nV_{nAB}, \nonumber \\
 \hat{\check\Theta}_{AB} &\equiv& \check\Theta_{AB}+v^n\check V_{nAB}.
\end{eqnarray}
Let us compute these transformation parameters for our ellipsoid background.
First of all, our condition (\ref{wr0}) on the Killing spinor guarantees
that $w=\Theta=0$. Moreover one can show
\begin{equation}
 L_{ab}\equiv0,\quad
 \hat\Theta^A_{\;~B}= \Big(-\frac1{2\ell}-\frac1{2\tilde\ell}\Big)\cdot
 (\pauli^3)^A_{~B}
\end{equation}
using the explicit form of vielbein, spin connection and the background
$SU(2)_\text{R}$ gauge field obtained above. It follows that our Killing
spinor is invariant under $\susy^2$.
\begin{eqnarray}
&& \susy^2\xi_A = i{\cal L}_v\xi_A-\xi_B\hat\Theta^B_{\;~A}=0,
 \nonumber \\
&& \susy^2\bar\xi_A = i{\cal L}_v\bar\xi_A-\bar\xi_B\hat\Theta^B_{\;~A} = 0.
\end{eqnarray}
The background fields $V_{m~B}^{~A}, T_{kl},\bar T_{kl}, M$ are also
invariant under $\susy^2$ since they are constructed from
${\cal L}_v$-invariant functions and Killing spinor.

To determine the action of $\susy^2$ on all the fields, we still need to
determine $\check\xi_A,\bar{\check\xi}_A$ and the background
$SU(2)_{\check{\text{R}}}$ gauge field $\check V_{m~A}^{~B}$ which have
been left somewhat ambiguous. Hereafter we take the following solution
of (\ref{checkxi}).
\begin{equation}
 \check\xi_A= \cot\tfrac\rho2\xi_A,\quad
 \bar{\check\xi}_A= -\tan\tfrac\rho2\bar\xi_A.
\label{checkxi2}
\end{equation}
Note that this has an effect of gauge fixing the local
$SU(2)_{\check{\text{R}}}$ symmetry relative to $SU(2)_\text{R}$, and the
following choice of the $SU(2)_{\check{\text{R}}}$ gauge field is
consistent with it.
\begin{equation}
 \check V_{m~A}^{~B} ~=~ V_{m~A}^{~B}\,.
\end{equation}
Using (\ref{checkxi2}) one can also show
\begin{equation}
 \check\Theta_{AB}=\Theta_{AB},~~\text{therefore}~~
 \hat{\check\Theta}_{AB}=\hat\Theta_{AB},
\end{equation}
and conclude that all the background fields are invariant under $\susy^2$.

\paragraph{Omega-background revisited.}

Here we focus on the behavior of our ellipsoid background near 
the north and south poles.

Near the north pole where $x_0\simeq r$ in (\ref{ellipsoids}), the other
four coordinates $(x_1,\cdots,x_4)$ can be regarded as the Cartesian
coordinates on $\mathbb R^4$. The norm of $\bar{\bs\xi}$ approaches a
constant while that of $\bs\xi$ is proportional to the radial distance
from the pole. In a suitable gauge, the Killing spinor should therefore
take the form
\begin{equation}
 \bar\xi^{\dot\alpha}_{~A}\simeq
 \frac1{\sqrt2}\delta^{\dot\alpha}_{~A},\quad
 \xi_{\alpha A}\simeq
-\frac1{2\sqrt2\ell}(x_1\sigma_2-x_2\sigma_1)_{\alpha A}
-\frac1{2\sqrt2\tilde\ell}(x_3\sigma_4-x_4\sigma_3)_{\alpha A}
\label{KSnp}
\end{equation}
so that
\begin{eqnarray}
  2\bar\xi^A\bar\sigma^m\xi_A\cdot
  \frac\partial{\partial x_m}
 &=&
  \frac 1\ell\Big(
   x_1\frac\partial{\partial x_2}
  -x_2\frac\partial{\partial x_1}\Big)
 +\frac 1{\tilde\ell}\Big(
   x_3\frac\partial{\partial x_4}
  -x_4\frac\partial{\partial x_3}\Big)
 \nonumber \\ &=&
  \frac 1\ell\partial_\varphi+\frac 1{\tilde\ell}\partial_\chi.
\end{eqnarray}
The first equation in (\ref{KSnp}) indicates that near the north
pole our supersymmetry approach that of the topologically twisted
theory which identifies the dotted spin $SU(2)$ index with the $SU(2)$
R-symmetry index. From this viewpoint, the second equation in
(\ref{KSnp}) tells nothing but the fact that $\ell^{-1},\tilde\ell^{-1}$
play the role of the Omega-deformation parameters $\epsilon_1,\epsilon_2$
\cite{Nekrasov-O}. Note that, for the spinor field (\ref{KSnp}) to
satisfy Killing spinor equation (\ref{ks1}) and (\ref{ks2}) on flat
$\mathbb R^4$, one has to turn on the background field as follows,
\begin{eqnarray}
&&
 T^{\Omega}\equiv\frac12 T^{\Omega}_{mn}\td x^m\td x^n =
 \frac1{16}\Big(\frac1{\tilde\ell}-\frac1\ell\Big)
 \big(\td x_1\td x_2-\td x_3\td x_4\big),
 \nonumber \\
&&
V^{\Omega}=\bar T^{\Omega}=M^{\Omega}=0.
\label{Omega}
\end{eqnarray}
In other words, Omega background with $\epsilon_1\ne\epsilon_2$ is
related to a flat $\mathbb R^4$ with constant background field $T_{kl}$.

In much the same way, near the south pole one can choose a gauge in
which $\xi_\alpha^A$ is proportional to the identity matrix. There the
supersymmetry approaches that of the anti-topologically twisted theory
with Omega deformation. One can also relate the flat $\mathbb R^4$ with
constant $\bar T_{mn}$ to Omega background of the anti-topologically
twisted theory.

It remains to check whether our ellipsoid background is regular at the two
poles. To do this, we rewrite the above regular Omega background
(\ref{Omega}) with the following polar coordinates of $\mathbb R^4$,
\begin{equation}
\begin{array}{rcl}
 x_1 &=& \ell \rho\cos\theta\cos\varphi, \\
 x_2 &=& \ell \rho\cos\theta\sin\varphi,
\end{array}
\quad
\begin{array}{rcl}
 x_3 &=& \tilde\ell \rho\sin\theta\cos\chi, \\
 x_4 &=& \tilde\ell \rho\sin\theta\sin\chi.
\end{array}
\end{equation}
The auxiliary field $T$ for the Omega background then takes the form,
\begin{equation}
T^{\Omega} ~=~ \frac1{16f}\left(\frac1{\tilde\ell}-\frac1\ell\right)
 \Big\{\ell\sin\theta(E^1E^3+E^2E^4)
      -\tilde\ell\cos\theta(E^1E^4-E^2E^3)\Big\},
\end{equation}
where $E^a$ are the natural vielbein one-forms on $\mathbb R^4$ in the
polar frame,
\begin{equation}
 E^1=\rho e^1,\quad
 E^2=\rho e^2,\quad
 E^3=\rho e^3 + h_{(0)}\td\rho,\quad
 E^4=g_{(0)}\td\rho.
\end{equation}
Here $h_{(0)}$ and $g_{(0)}$ denote the values of the functions $h$ and
$g$ in (\ref{fgh}) at $\rho=0$.
Then one finds
\begin{equation}
 {\bf T}^{\Omega}~\equiv~ -iT^{\Omega}_{mn}\sigma^{mn}
 ~=~ \frac14\Big(\frac1f-\frac1{g_{(0)}}\Big)\pauli^1_\theta
 +\frac{h_{(0)}}{4fg_{(0)}}\pauli^2_\theta,
\end{equation}
which agrees with our special solution (\ref{solsp}) near
the north pole. However, there is a finite mismatch between the value of
$\bar T$, which is zero on the Omega background (\ref{Omega}) but
nonvanishing near the north pole of (\ref{solsp}). This indicates that
our special solution has singularity at the two poles and a suitable
nonzero $c_1,c_2$ has to be chosen so as to cancel it.
A simple choice which leads to $\bar T=0$ at the north pole and $T=0$
at the south pole is given by
\begin{equation}
 c_1=\frac18\Big(\frac1f-\frac1g\Big)\sin\rho\cos\rho,\quad
 c_2=\frac h{8fg}\sin\rho\cos\rho.
\end{equation}
One is still left with the freedom to shift the $c$'s by functions which
vanish as $\sin^2\rho$ or faster near the two poles.

\section{Explicit Path Integration}\label{sec:localization}

Here we use the SUSY localization principle and evaluate partition
functions of general SW theories on the ellipsoid backgrounds. Our analysis
follows closely that of \cite{Pestun}. We first focus on the theories
with vector multiplets only, and introduce matter hypermultiplets later.

\paragraph{Saddle points for SYM theories.}

According to the SUSY localization principle, non-zero contribution
to the path integral arises only from saddle points which are
characterized by
\[
 \susy \Psi ~=~ 0\quad
 \text{for all the fermions $\Psi$}.
\]
The first step in computing partition function is to find out the
saddle point locus. Though we have to modify the supercharge $\susy$
upon introducing BRST ghost system, the saddle point locus remain the
same.

To find out the saddle point locus for vector multiplets, it is
convenient to study the following quantity,
\begin{equation}
{\cal I}_\text{vec}~\equiv~
\text{Tr}\Big[
 (\susy\lambda_{\alpha A})^\dagger
 (\susy\lambda_{\alpha A})
+(\susy\bar\lambda^{\dot\alpha}_{~A})^\dagger
 (\susy\bar\lambda^{\dot\alpha}_{~A})
 \Big],
\end{equation}
which is manifestly positive semi-definite and vanishes on saddle
points. Using the transformation law and the reality condition
(\ref{realityvec}), one can rewrite it as follows,
\begin{eqnarray}
{\cal I}_\text{vec} &=&
 \text{Tr}\bigg[
  D_m\phi_1D^m\phi_1-[\phi_1,\phi_2]^2
 -\frac12(D_{AB}+i\phi_1w_{AB})(D^{AB}+i\phi_1w^{AB})
 \nonumber \\ &&\hskip5mm
 +\,\xi^A\xi_A\Big(F_{mn}^--4\phi_2T_{mn}-4\phi_2S_{mn}
            +\frac1{\xi^A\xi_A}v_{[m}D_{n]^-}\phi_2\Big)^2
 \nonumber \\ &&\hskip5mm
 +\,\bar\xi_A\bar\xi^A\Big(F_{mn}^++4\phi_2\bar T_{mn}+4\phi_2\bar S_{mn}
            -\frac1{\bar\xi_A\bar\xi^A}v_{[m}D_{n]^+}\phi_2\Big)^2
 \nonumber \\ &&\hskip5mm
 +\,\frac1{4\xi^A\xi_A\cdot\bar\xi_B\bar\xi^B}\big(v^mD_m\phi_2\big)^2
 \bigg],
\label{Ivec}
\end{eqnarray}
where the suffix $\pm$ for antisymmetric tensors indicates the self-dual
or anti-self-dual parts, and we introduced
\begin{eqnarray}
&& \phi_1 \equiv i(\phi+\bar\phi),\quad
   \phi_2 \equiv \phi-\bar\phi,
 \nonumber \\
&& w_{AB} \equiv
    \frac{4\xi_A\sigma^{mn}\xi_B\,(T_{mn}-S_{mn})}{\xi^C\xi_C}
 = -\frac{4\bar\xi_A\bar\sigma^{mn}\bar\xi_B\,(\bar T_{mn}-\bar S_{mn})}
         {\bar\xi_C\bar\xi^C}\,.
\end{eqnarray}
Note that $w_{AB}$ here satisfies the condition (\ref{FIcoeff}) and
therefore can be used to construct FI Lagrangian.

The saddle point condition for $\phi_2$ and $A_m$ is to be derived from
the last three terms in the r.h.s. of (\ref{Ivec}). We argue that it is
given by
\begin{equation}
 \phi_2=A_m=0\quad\text{up to gauge choice.}
\label{saddle1}
\end{equation}
For round sphere with $T_{mn}\equiv\bar T_{mn}\equiv0$, one finds that
the last three terms can be reorganized into a different ``sum of squares''
up to total derivatives,
\begin{equation}
{\cal I}_\text{vec}~=~\text{Tr}\bigg[
 \cdots+(D_m\phi_2)^2
 +\xi^A\xi_A(F^-_{mn}+4\phi_2S_{mn})^2
 +\bar\xi_A\bar\xi^A(F^+_{mn}-4\phi_2\bar S_{mn})^2
\bigg].
\label{Ivec2}
\end{equation}
This gives a much simpler saddle point condition which immediately leads
to (\ref{saddle1}) when combined with Bianchi identity
$D_{[l}F_{mn]}=0$. However, as soon as the sphere is deformed, this
reorganizing is no longer possible and one has to deal with more
complicated saddle point condition which follows from (\ref{Ivec}). But
if there are nontrivial solutions to the original saddle point condition
on some deformed sphere, they should be continuously connected to
nontrivial solutions on round sphere. Such solutions would have to be
singular, since they do not minimize ${\cal I}_\text{vec}$ of
(\ref{Ivec2}) which differs from (\ref{Ivec}) only by total
derivatives. Thus we believe that (\ref{saddle1}) is the only solution
to the saddle-point condition. It would be nice to prove this claim
rigorously, though we will base our subsequent analysis on this claim
and obtain the most natural generalization of the result for round sphere.

Once (\ref{saddle1}) is settled, then the condition for the remaining
fields are easily solved. The saddle points are thus labeled by a Lie
algebra valued constant $a_0$, and are given by the equations
\begin{equation}
 A_m=0,\quad
 \phi=\bar\phi=-\frac i2a_0\,,\quad
 D_{AB}= -ia_0w_{AB}\,.
\end{equation}
The values of super-Yang-Mills action (\ref{LYM}) and FI term (\ref{LFI})
on this saddle point are
\begin{eqnarray}
 \frac1{g_\text{YM}^2}
 \int \td^4x\sqrt g{\cal L}_\text{YM}\Big|_\text{saddle point} &=&
 \frac{8\pi^2}{g_\text{YM}^2}\,\ell\tilde\ell\text{Tr}(a_0^2)\,,
 \nonumber \\
 \zeta
 \int \td^4x\sqrt g{\cal L}_\text{FI}\Big|_\text{saddle point} &=&
 -16i\pi^2\ell\tilde\ell\zeta a_0\,.
\end{eqnarray}
They are independent of the precise choices of $c_1,c_2,c_3$ as long as
they are smooth.

\paragraph{Ghosts and BRST symmetry.}

For gauge fixing, we proceed in the same way as \cite{Pestun}. Let us
introduce the Faddeev-Popov ghost field $c$ and define the BRST
transformation by
\begin{equation}
\begin{array}{rcl}
 \brs A_m &=& D_mc, \\
 \brs \phi &=& i[c,\phi], \\
 \brs\bar\phi &=& i[c,\bar\phi],
\end{array}
\quad
\begin{array}{rcl}
 \brs \lambda_A &=& i\{c,\lambda_A\}, \\
 \brs\bar\lambda_A &=& i\{c,\bar\lambda_A\}, \\
 \brs D_{AB} &=& i[c,D_{AB}].
\end{array}
\end{equation}
We require the square of $\brs$ to be a constant gauge rotation with
parameter $a_0$, so we set
\begin{equation}
 \brs c ~=~ icc+a_0.
\end{equation}
The sum of the SUSY and the BRST transformations,
$\qtot\equiv\susy+\brs$, will be the relevant fermionic symmetry in the
application of localization principle later on. Requiring its square to
act on all the fields as
\begin{equation}
 \qtot^2 ~=~ i{\cal L}_v + \text{Gauge}(a_0)+\text{R}_{SU(2)}(\hat\Theta_{AB}),
\end{equation}
one finds that the supersymmetry transformation of $c$ has to be,
\begin{equation}
 \susy c ~=~ -\hat\Phi ~=~ -\phi_1-i\cos\rho\phi_2+iv^nA_n.
\end{equation}
One also finds that the constant variable $a_0$ has to be invariant,
\begin{equation}
 \susy a_0 ~=~ \brs a_0 ~=~ 0.
\end{equation}
We furthermore introduce the antighost multiplet with the transformation
rules,
\begin{equation}
\begin{array}{rcl}
 \brs \bar c &=& B, \\
 \susy \bar c &=& 0,
\end{array}
\quad
\begin{array}{rcl}
 \brs B &=& i[a_0,\bar c], \\
 \susy B &=& i{\cal L}_v\bar c,
\end{array}
\end{equation}
and the multiplets of constant fields which will be used to freeze the
constant modes of $c$ and $\bar c$.
\begin{equation}
\begin{array}{rcl}
 \brs \bar a_0 &=& \bar c_0, \\
 \susy\bar a_0 &=& 0,
\end{array}
\quad
\begin{array}{rcl}
 \brs \bar c_0 &=& i[a_0,\bar a_0], \\
 \susy\bar c_0 &=& 0,
\end{array}
\qquad
\begin{array}{rcl}
 \brs  B_0 &=& c_0, \\
 \susy B_0 &=& 0,
\end{array}
\quad
\begin{array}{rcl}
 \brs  c_0 &=& i[a_0,B_0], \\
 \susy c_0 &=& 0.
\end{array}
\end{equation}

To fix a gauge correctly, the standard way is to choose a set of
conditions $G[A_m,\phi,\cdots]$ and shift the Lagrangian by the
gauge-fixing term
\begin{equation}
{\cal L}_\text{GF} = \brs{\cal V}_\text{GF},
\quad
{\cal V}_\text{GF}\equiv\text{Tr}\left(\bar cG+\bar cB_0+c\bar a_0\right).
\label{LGF}
\end{equation}
We will later find it convenient to choose
\begin{equation}
 G ~=~ i\partial_mA^m+i{\cal L}_v(\cos\rho\phi_2-v^mA_m).
\end{equation}
For the computation of partition function using localization principle,
it is more convenient to replace $\brs$ in (\ref{LGF}) by
$\qtot=\susy+\brs$. As explained in \cite{Pestun} this replacement does not change
the value of partition function.

Now that the gauge-fixed system has the fermionic symmetry
$\qtot\equiv\susy+\brs$, we need to revisit the condition for the saddle
points
\begin{equation}
 \qtot\Psi ~=~ \susy\Psi+\brs\Psi~=~0~~\text{for all the fermions}~\Psi.
\end{equation}
For the fermions in vector multiplets, the added term $\brs \Psi$ is always
bilinear in fermions so that the condition for saddle points does not
change. For the ghost $c$, the saddle point condition gives
\begin{equation}
 \qtot c ~=~ icc+a_0-\phi_1-i\cos\rho\phi_2+iv^nA_n~=~0.
\end{equation}
Thus $a_0$ is to be identified with the constant value of $\phi_1$ at
saddle points.

\paragraph{One-loop determinant.}

The value of path integral does not change under the shifts of the
original Lagrangian by any $\qtot$-exact quantities,
${\cal L}\to {\cal L}+t\qtot{\cal V}$. We take the regulator
$\qtot{\cal V}$ so that its bosonic part is positive definite and is
strictly positive anywhere away from saddle points. Since $t$ can be
taken arbitrarily large, Gaussian approximation is exact for the
path integration over the fluctuations away from saddle points.

We begin by introducing some new notations for later convenience.
\begin{eqnarray}
 \Psi &\equiv& \susy\phi_2~=~
 -i\xi^A\lambda_A-i\bar\xi^A\bar\lambda_A,
 \nonumber \\
 \Psi_m &\equiv& \susy A_m~=~
 i\xi^A\sigma_m\bar\lambda_A-i\bar\xi^A\bar\sigma_m\lambda_A,
 \nonumber \\
 \Xi_{AB} &\equiv& 2\bar\xi_{(A}\bar\lambda_{B)}-2\xi_{(A}\lambda_{B)}.
\end{eqnarray}
The inverse of this relation is
\begin{eqnarray}
 \lambda_A &=&
  +i\xi_A\Psi
  -i\sigma^m\bar\xi_A\Psi_m
  +\xi^B\Xi_{BA},
 \nonumber \\
 \bar\lambda_A &=&
 -i\bar\xi_A\Psi
 -i\bar\sigma^m\xi_A\Psi_m
  +\bar\xi^B\Xi_{BA}.
\label{lambdapsi}
\end{eqnarray}
As the regulator, we take the $\qtot$-transform of the following
quantity which has manifestly positive semi-definite bosonic part
${\cal I}_\text{vec}$,
\begin{equation}
 {\cal V}~=~
 \text{Tr}\Big[
 (\qtot\lambda_{\alpha A})^\dagger\lambda_{\alpha A}
 +(\qtot\bar\lambda_A^{\dot\alpha})^\dagger\bar\lambda_A^{\dot\alpha}\Big].
\end{equation}
Inserting (\ref{lambdapsi}) into this and combining with the gauge
fixing term, one finds
\begin{equation}
 {\cal V}+{\cal V}_\text{GF}~=~
 \text{Tr}\Big[
     (\qtot\Psi)^\dagger\Psi
    +(\qtot\Psi^m)^\dagger\Psi_m
    +\tfrac12(\qtot\Xi_{AB})^\dagger\Xi_{AB}
    +\bar cG+\bar cB_0+c\bar a_0
\Big].
\end{equation}

The integration over all the variables except for the constant $a_0$
will be carried out under the (exact) Gaussian approximation, with the
weight given by $\qtot({\cal V}+{\cal V}_\text{GF})$ truncated up to
quadratic order. In doing this, we move to a new set of path integration
variables which consists of
\begin{equation}
 X \equiv (\phi_2,A_m;\bar a_0,B_0), \quad
 \Xi \equiv (\Xi_{AB},\bar c,c)
\end{equation}
and their superpartners $\qtot X,\qtot\Xi$.
In terms of these variables one can write
\begin{equation}
 {\cal V}+{\cal V}_\text{GF}\Big|_\text{quad.} ~=~
 (\qtot X, \Xi)\left( \begin{array}{cc}
 D_{00} & D_{01} \\ D_{10} & D_{11} \end{array}\right)
 \left(\begin{array}{c}
 X \\ \qtot\Xi \end{array}\right).
\end{equation}
The Gaussian integration gives the square root of the ratio of
determinants of kinetic operators for boson and fermions.
Using the fact that the operators $D_{ij}$ commute with
$\qq\equiv\qtot^2$, one finds after some algebra that
\begin{equation}
 \frac{\text{det}K_\text{fermion}} {\text{det}K_\text{boson}} ~=~
 \frac{\text{det}_\Xi\qq}{\text{det}_X\qq}~=~
 \frac{\text{det}_{\text{Coker}D_{10}}\qq}
      {\text{det}_{\text{Ker}D_{10}}\qq}.
\end{equation}
Thus the ratio of determinants can be determined from the spectrum of
the operator $\qq$ on the kernel and cokernel of a differential operator $D_{10}$,
which is encoded in the index
\begin{equation}
 \text{ind}D_{10} ~\equiv~
 \text{Tr}_{\text{Ker}D_{10}}\big(e^{-i\qq t}\big)
-\text{Tr}_{\text{Coker}D_{10}}\big(e^{-i\qq t}\big).
\end{equation}

\paragraph{Index of transversally elliptic operators.}

In computing this index, we first drop the terms containing constant
fields $B_0,\bar a_0$ from ${\cal V}_\text{GF}$. These constant fields
are thus regarded as sitting in the kernel of $D_{10}$ and making a
contribution 2 to the index. To obtain the remaining contribution, we
read off the differential operator $D_{10}$ from
\begin{equation}
 \Xi D_{10} X + \Xi D_{11}\qtot\Xi
 ~=~ \text{Tr}\left[
  \bar cG-D_mc(\qtot\Psi^m)^\dagger+\tfrac12\Xi_{AB}(\qtot\Xi_{AB})^\dagger
 \right]\Big|_\text{quad},
\end{equation}
where we have, up to non-linear terms,
\begin{eqnarray}
 (\qtot\Psi_m)^\dagger &=&
 -i{\cal L}_vA_m+D_m(\hat\Phi-2i\cos\rho\,\phi_2+2iv^nA_n),
 \nonumber \\
 (\qtot\Xi_{AB})^\dagger &=&
 -\xi^A\sigma^{kl}\xi^B
 (F_{kl}-8\phi T_{kl}+8\bar\phi S_{kl})
 \nonumber \\ &&
 +\bar\xi^A\bar\sigma^{kl}\bar\xi^B
 (F_{kl}-8\bar\phi\bar T_{kl}+8\phi\bar S_{kl})
 -4\xi^{(A}\sigma^n\bar\xi^{B)}D_n\phi_2 -D^{AB}.
\end{eqnarray}
It turns out that the operator $D_{10}$ is not elliptic but transversally
elliptic with respect to the isometry ${\cal L}_v$ of the ellipsoid.
Let us show this by computing explicitly its symbol.

We identify the fields $X$ and $\Xi$ with sections of bundles
$E_0$ and $E_1$ over the ellipsoid ${\cal X}$, and therefore
$D_{10}:\Gamma(E_0)\to \Gamma(E_1)$. Its symbol $\sigma(D_{10})$ is then
obtained by retaining only the terms with highest order of derivatives
and making the replacement $\partial_{x^i}\to ip_i$. Thus
$\sigma(D_{10})$ is a homomorphism between two vector bundles
$\pi^\ast E_0,\pi^\ast E_1$ over the cotangent bundle
$\pi:T^\ast{\cal X}\to {\cal X}$. The index of transversally elliptic
operators is known to be uniquely determined by their symbols.

To write the symbol explicitly, it is convenient to introduce four unit
vector fields $u_a^m~(a=1,\cdots,4)$ by the formula
\begin{eqnarray}
 -2i(\pauli^a)^A_{~\;B}\bar\xi^B\bar\sigma^m\xi_A
 &=& \sin\rho\;u^m_a~~(a=1,2,3),
 \nonumber \\
  2  \bar\xi^A\bar\sigma^m\xi_A
 &=& \sin\rho\;u^m_4
\end{eqnarray}
and parametrize the momenta in the local orthonormal frame defined by
the vielbein $u^m_a$. For example, by a slight abuse of the notation,
one can write
\begin{eqnarray}
 4\bar\xi^A\bar\sigma^m\xi_B\,\partial_m
 &=& \sin\rho\,(\bar\sigma^a)^A_{~B}\;u_a^m\partial_m,
 \nonumber \\
 -4\xi^A\sigma^m\bar\xi_B\,\partial_m
 &=& \sin\rho\,(\sigma^a)^A_{~B}\;u_a^m\partial_m,
\label{sigmaaAB}
\end{eqnarray}
and in particular
\begin{equation}
{\cal L}_v\equiv v^m\partial_m = \sin\rho u_4^m\cdot ip_m
 = i\sin\rho\cdot p_4.
\end{equation}
Using this notation together with
$\Xi_a\equiv\frac12\Xi^A_{~\,B}(\pauli_a)^B_{~A}$, one finds
\begin{equation}
 \Xi\,\sigma(D_{10})X  =
 (\Xi_1,\Xi_2,\Xi_3,-\bar c,ic)
 \left(\begin{array}{ccccc}
  c_\rho p_4 & p_3 & -p_2 & -c_\rho p_1 & -s_\rho p_1 \\
  -p_3 & c_\rho p_4 & p_1 & -c_\rho p_2 & -s_\rho p_2 \\
  p_2 & -p_1 & c_\rho p_4 & -c_\rho p_3 & -s_\rho p_3 \\
  p_1 &  p_2 & p_3 & c_\rho^2 p_4 & c_\rho s_\rho p_4 \\
  p_1p_4 & p_2p_4 & p_3p_4 & p_4^2-2s_\rho p_a^2 & 2c_\rho p_a^2
 \end{array}\right)
 \left(\begin{array}{c}
 A_1 \\ A_2 \\ A_3 \\ A_4 \\ \phi_2\end{array}\right),
\end{equation}
where we denoted $s_\rho\equiv\sin\rho,\,c_\rho\equiv\cos\rho$. The
$5\times 5$ matrix in the middle can be block diagonalized by a suitable
change of variables within $X$ and $\Xi$.
\begin{equation}
 \sigma(D_{10})=
 \left(\begin{array}{ccccc}
 1 &&&& \\ &1&&&\\ &&1&&\\ &&&1& \\ &&&p_4&1
 \end{array}\right)
 \left(\begin{array}{ccccc}
  c_\rho p_4 & p_3 & -p_2 & -p_1 & \\
  -p_3 & c_\rho p_4 & p_1 & -p_2 & \\
  p_2 & -p_1 & c_\rho p_4 & -p_3 & \\
  p_1 &  p_2 & p_3 & c_\rho  p_4 & \\
     &  &  &  & -2p_a^2+s_\rho p_4^2
 \end{array}\right)
 \left(\begin{array}{ccccc}
 1&&&&\\ &1&&&\\ &&1&&\\ &&&c_\rho&s_\rho\\ &&& s_\rho&-c_\rho
 \end{array}\right).
\end{equation}
The lower-right $1\times 1$ block should give a trivial contribution to the
index, since the corresponding differential operator should have just
one-dimensional kernel and cokernel of constant functions. So the nontrivial
contribution to the index arises from the upper-left $4\times4$ block of
the matrix in the middle,
\begin{equation}
 \sigma(D_{10}')~=~
 \left(\begin{array}{cccc}
  c_\rho p_4 & p_3 & -p_2 & -p_1 \\
  -p_3 & c_\rho p_4 & p_1 & -p_2 \\
  p_2 & -p_1 & c_\rho p_4 & -p_3 \\
  p_1 &  p_2 & p_3 & c_\rho  p_4
 \end{array}\right)\,.
\end{equation}
Near the two poles, the symbol is that of the standard self-dual or
anti-self-dual complex on $\mathbb R^4$,
\begin{eqnarray}
 (\cos\rho=+1) &&
 \Omega^0\stackrel{{\rm d}}\longrightarrow\Omega^1
 \stackrel{{\rm d}^+}\longrightarrow\Omega^{2+},
 \nonumber \\
 (\cos\rho=-1) &&
 \Omega^0\stackrel{{\rm d}}\longrightarrow\Omega^1
 \stackrel{{\rm d}^-}\longrightarrow\Omega^{2-}.
\end{eqnarray}

A differential operator is called elliptic if its symbol is invertible
for nonzero $p_a$. The above symbol $\sigma$ is not invertible at the
equator $\cos\rho=0$ since
$\sigma\sigma^T=  (p_1^2+p_2^2+p_3^2+\cos^2\rho p_4^2)\cdot\text{id}$
as one can easily check. But if we restrict the momentum to be
orthogonal to the vector $v$, namely $p_4\equiv0$, then $\sigma$ is
invertible as long as $(p_1,p_2,p_3)$ are not all zero. The
corresponding differential operator is then called transversally
elliptic with respect to the symmetry ${\cal L}_v$. The kernel and
cokernel of transversally elliptic operators are generally infinite
dimensional, though they are both decomposed into finite dimensional
eigenspaces of $\qq$. Therefore, there is a bit more difficulty in the
computation of index for transversally elliptic operators as compared to
elliptic ones.

The operator $e^{-i\qq t}$ is a combination of a finite rotation of the
ellipsoid, gauge rotation and $SU(2)_\text{R}$ rotation. Its action on
an adjoint-valued field ${\cal O}$ takes the form
\begin{equation}
 e^{-i\qq t}{\cal O}(x^m)~=~
 \gamma_{[{\cal O}]}\cdot e^{a_0t}{\cal O}(\tilde x^m)e^{-a_0t},
 \quad
 \Big(\tilde\varphi=\varphi+\tfrac t\ell,~
      \tilde\chi=\chi+\tfrac t{\tilde\ell}\Big)
\end{equation}
where the coefficient $\gamma_{[{\cal O}]}$ encodes the action on the
vector and $SU(2)_\text{R}$ indices of the field ${\cal O}$. For
simplicity, let us temporarily take the gauge group to be abelian.

Regarding the index as the difference of the trace of $e^{-i\qq t}$
over $\Gamma(E_0)$ and $\Gamma(E_1)$, it should be written as a sum of
contributions from the two fixed points where $\tilde x^m=x^m$. 
According to the Atiyah-Bott formula, the index is given by
\begin{equation}
 \text{ind}(D_{10}')~=~ \sum_{x:\text{fixed point}}
 \frac{\text{Tr}_{E_0}(\gamma)-\text{Tr}_{E_1}(\gamma)}
      {\text{det}(1-\partial\tilde x/\partial x)}\,,
\end{equation}
where the determinant factor is understood to arise from
$\td^4x\delta^4(\tilde x(x)-x)$.
Near the north pole, the operator $e^{-i\qq t}$ acts on the local
coordinates $z_1=x_1+ix_2,z_2=x_3+ix_4$ as
\begin{equation}
 \tilde z_1= e^{\frac{it}{\ell}}z_1\equiv q_1z_1,\quad
 \tilde z_2= e^{\frac{it}{\tilde\ell}}z_2\equiv q_2z_2.
\end{equation}
Therefore
\begin{equation}
 \text{det}(1-\partial\tilde x/\partial x)~=~
 (1-q_1)(1-\bar q_1)(1-q_2)(1-\bar q_2),
\end{equation}
where $q_1\bar q_1=q_2\bar q_2=1$.
The value of $\gamma$ for various fields reads
\begin{equation}
\begin{array}{rcl}
 \gamma[A_{z_1}]&=&q_1,\\
 \gamma[A_{z_2}]&=&q_2,\\
 \gamma[A_{\bar z_1}]&=&\bar q_1,\\
 \gamma[A_{\bar z_2}]&=&\bar q_2,
\end{array}
\quad
\begin{array}{rcl}
 \gamma[\Xi_{11}] &=& \bar q_1\bar q_2, \\
 \gamma[\Xi_{12}] &=& 1, \\
 \gamma[\Xi_{22}] &=& q_1q_2, \\
 \gamma[\bar c] &=&1.
\end{array}
\end{equation}
These are enough to compute the contribution from the north pole.
Combining it with the similar contribution from the south pole and
2 from constant modes, one obtains
\begin{equation}
 \text{ind}(D_{10})~=~
 \left[-\frac{1+q_1q_2}{(1-q_1)(1-q_2)}\right]
+\left[-\frac{1+q_1q_2}{(1-q_1)(1-q_2)}\right]
+2\,.
\end{equation}

To extract the information on the multiplicity of eigenvalues of $\qq$,
one needs to expand this expression into power series in $q_1,q_2$.
The expansion does not seem to be unique, and the correct way should be
found by investigating a suitable deformation of the symbol to make it
non-degenerate everywhere away from the two poles. As was explained in
\cite{Atiyah} and reviewed in \cite{Pestun}, this is the main point of
difficulty in computing the index of transversally elliptic operators.
At the end of the day, the correct way is to expand the first term in
positive series and the second term in negative series. Thus we arrive
at
\begin{equation}
 \text{ind}(D_{10})~=~
 2-\sum_{m,n\ge0}\Big(
  q_1^{m}q_2^{n}
 +q_1^{m+1}q_2^{n+1}
 +q_1^{-m}q_2^{-n}
 +q_1^{-m-1}q_2^{-n-1}
 \Big)\,.
\end{equation}
For non-abelian gauge group $G$, we take $a_0$ to be in the Cartan
subalgebra. Then each term in the above is multiplied by
\begin{equation}
 \text{rk}G + \sum_{\alpha\in\Delta}\exp\left(ta_0\cdot\alpha\right)
\end{equation}
where the sum runs over all roots. This finishes the computation of the
index $\text{ind}(D_{10})$.

\paragraph{Infinite-product formula.}

The one-loop determinant can be easily computed by extracting the
spectrum of eigenvalues of $\qq$ from the index. Up to normalization factors
depending only on $\ell$ and $\tilde\ell$, it is given by
\begin{eqnarray}
 Z_\text{1-loop}^\text{vec} &=&
 \left[\frac{\text{det}K_\text{fermion}}{\text{det}K_\text{boson}}
 \right]^{\frac12}
 \nonumber \\
 &=& \prod_{\alpha\in\Delta_+}\frac{1}{(\hat a_0\cdot\alpha)^2}\prod_{m,n\ge0}
 \big(mb+nb^{-1}+Q+i\hat a_0\cdot\alpha\big)
 \big(mb+nb^{-1}+i\hat a_0\cdot\alpha\big)
 \nonumber \\ && \hskip33mm\cdot
 \big(mb+nb^{-1}+Q-i\hat a_0\cdot\alpha\big)
 \big(mb+nb^{-1}-i\hat a_0\cdot\alpha\big)
 \nonumber \\ &=&
 \prod_{\alpha\in\Delta_+}\frac
 {\Upsilon(i\hat a_0\cdot\alpha)
  \Upsilon(-i\hat a_0\cdot\alpha)}
 {(\hat a_0\cdot\alpha)^2} \;,
\label{Zvec1lp}
\end{eqnarray}
where we introduced $b\equiv (\ell/\tilde\ell)^{1/2}, Q\equiv b+\frac1b$
and $\hat a_0\equiv\sqrt{\ell\tilde\ell}a_0$. We also used the function
$\Upsilon(x)$ which has zeroes at
$x=Q+mb+\frac nb,-mb-\frac nb\;(m,n\in\mathbb Z_{\ge0})$
to express the appropriately regularized infinite products. It is
characterized by
\begin{equation}
\Upsilon(x) = \Upsilon(Q-x),\quad
\Upsilon(Q/2) = 1,
\end{equation}
as well as the shift relations
\begin{eqnarray}
\Upsilon(x+b) &=& \Upsilon(x)\gamma(bx)b^{1-2bx}, \nonumber \\
\Upsilon(x+\tfrac1b) &=& \Upsilon(x)\gamma(x/b)b^{\frac{2x}b-1}. \quad
\Big(\gamma(x)\equiv\Gamma(x)/\Gamma(1-x)\Big)
\end{eqnarray}
The function $\Upsilon(x)$ was used in \cite{ZZ} to write down the
three-point structure constant in Liouville CFT with coupling $b$. Thus
our result suggests that the ellipsoid is the correct background for SW
theories to reproduce Liouville or Toda correlators for general value of
the coupling $b$.

The final expression for partition function involves an integral with
respect to the saddle point parameter $a_0$ over the Lie algebra, but one
can restrict its integration domain to Cartan subalgebra. It gives
rise to the usual Vandermonde determinant factor which cancels with the
factor $(\hat a_0\cdot\alpha)^2$ in the denominator of (\ref{Zvec1lp}).

\paragraph{Inclusion of matter.}

Let us next study the case with hypermultiplet matters. The first thing
to do is to solve the saddle point condition. For round $S^4$ it was
shown in \cite{Pestun} that all the bosonic fields in hypermultiplets
have to vanish at saddle points; in other words there is no Higgs
branch. We claim this remains true on ellipsoids. The simplest way
to see this is to consider the zero locus of the bosonic part of
\begin{equation}
 {\hat{\cal L}}_\text{mat} ~=~ \qtot{\cal V}_\text{mat},
\label{Lreghyp}
\end{equation}
which is the same as the bosonic part of (\ref{Lmatful}). The auxiliary
field $F_A$ simply has to vanish. The scalar $q_A$ has mass term which
is smallest at the origin of the Coulomb branch where
$\phi=\bar\phi=-\frac i2a_0=0$, and its value $\frac14(R+M)$ is
strictly positive anywhere on the ellipsoid at least when the deformation
from the round sphere with $T_{mn}=\bar T_{mn}=0$ is not large.

The one-loop determinant can be computed in the same way as for the
vector multiplets. We define new Grassmann-odd scalar fields which are
doublets under $SU(2)_\text{R}$ or $SU(2)_{\check{\text{R}}}$ by the formula,
\begin{equation}
 \Psi_A \equiv -i\xi_A\psi+i\bar\xi_A\bar\psi=\susy q_A,\quad
 \Xi_A \equiv \check\xi_A\psi-\bar{\check\xi}_A\bar\psi.
\end{equation}
The inverse of this is
\begin{equation}
 \psi=-2i\xi^A\Psi_A-2\check\xi^A\Xi_A,\quad
 \bar\psi=-2i\bar\xi^A\Psi_A-2\bar{\check\xi}^A\Xi_A.
\end{equation}
We then rewrite the regulator Lagrangian (\ref{Lreghyp}) truncated up to
quadratic order in terms of the variables $(q_A,\qtot q_A)$ and
$(\Xi_A,\qtot\Xi_A)$. The computation of the one-loop
determinant thus reduces to that of the index of an operator
$D_{10}^\text{mat}$ which can be read from the terms bilinear in
$\Xi_A$ and $q_A$ in ${\cal V}_\text{mat}$. Its symbol is given by
\begin{equation}
 \Xi_A[\sigma(D_{10}^\text{mat})]^A_{~B}q^B ~=~
 i\cos^2\tfrac\rho2\,\Xi_A(\sigma^ap_a)^A_{~B}q^B
-i\sin^2\tfrac\rho2\,\Xi_A(\bar\sigma^ap_a)^A_{~B}q^B\,,
\end{equation}
where we used the notation introduced in (\ref{sigmaaAB}). The
ellipticity of $D_{10}^\text{mat}$ is violated at $\rho=\frac\pi2$ but
it is transversally elliptic with respect to the isometry generated by
${\cal L}_v$.

Using Atiyah-Bott formula again, we compute the index from the action of
$\qq$ on fields at the two poles. At the north pole it is most
convenient to work with the Cartesian local coordinates
$x_1,\cdots,x_4$, in terms of which the metric is flat and the Killing
spinor takes the form (\ref{KSnp}). Here one can regard $q_A$ as
dotted spinor and identify $\Xi_A$ as undotted spinor $\psi$. Thus we find,
for example for $r$ free hypermultiplets,
\begin{eqnarray}
 \gamma[q^{A=1}_I]=q_1^{\frac12}q_2^{\frac12},&&
 \gamma[\psi_{\alpha=1}^I]=q_1^{\frac12}\bar q_2^{\frac12},
 \nonumber \\
 \gamma[q^{A=2}_I]=\bar q_1^{\frac12}\bar q_2^{\frac12},&&
 \gamma[\psi_{\alpha=2}^I]=\bar q_1^{\frac12}q_2^{\frac12}.
 \quad(I=1,\cdots,2r)
\end{eqnarray}
Combining the contribution from the two poles one finds the index
\begin{eqnarray}
 \text{ind}(D_{10}^\text{mat}) &=&
  2r\left[\frac{q_1^{1/2}q_2^{1/2}}{(1-q_1)(1-q_2)}\right]
 +2r\left[\frac{q_1^{1/2}q_2^{1/2}}{(1-q_1)(1-q_2)}\right]
 \nonumber \\ &=&
 2r\sum_{m,n\ge0}\Big(q_1^{m+\frac12}q_2^{n+\frac12}
                +q_1^{-m-\frac12}q_2^{-n-\frac12}\Big),
\end{eqnarray}
where we assumed that a regularization procedure similar to the case
with vector multiplet determines how to expand the first line into
power series.
For hypermultiplets coupled to gauge symmetry, the factor $2r$ is
replaced by a sum over the weight vectors of the corresponding
representation. For example, the hypermultiplet is said to be in a
representation $R$ of the gauge group if the index $I$ furnishes the
representation ${\cal R}=R\oplus\bar R$. Then the index is given by the
replacement
\begin{equation}
 2r ~\longrightarrow~
   \sum_{\rho\in{\cal R}}e^{ta_0\cdot\rho}
~=~\sum_{\rho\in R}\big(e^{ta_0\cdot\rho}+e^{-ta_0\cdot\rho}\big),
\end{equation}
where $\rho$ runs over all the weight vectors in a given representation
${\cal R}$ or $R$. This completes the computation of the index. It is
straightforward to translate this result into the matter one-loop
determinant,
\begin{eqnarray}
 Z_\text{1-loop}^\text{hyp}
 &=& \prod_{\rho\in R}\prod_{m,n\ge0}
 \big(mb+nb^{-1}+\tfrac Q2+i\hat a_0\cdot\rho\big)^{-1}
 \big(mb+nb^{-1}+\tfrac Q2-i\hat a_0\cdot\rho\big)^{-1}
 \nonumber \\ &=&
 \prod_{\rho\in R}\Upsilon(i\hat a_0\cdot\rho+\tfrac Q2)^{-1}.
\end{eqnarray}

\paragraph{Instanton contribution.}

In solving the saddle point condition for vector multiplet, the gauge
field was assumed to be smooth. Relaxing this assumption, one finds from
(\ref{Ivec}) that the gauge field strength can have nonzero
anti-self-dual components at the north pole where
$\xi^A\xi_A=\sin^2\frac\rho2=0$, or nonzero self-dual components at the
south pole where $\bar\xi_A\bar\xi^A=\cos^2\tfrac\rho2=0$.

The system near the north pole approaches the topologically twisted
theory with Omega deformation $\epsilon_1=\ell^{-1},\epsilon_2=\tilde\ell^{-1}$,
and the contribution of localized instantons is described by Nekrasov's
instanton partition function $Z_\text{inst}(a_0,\epsilon_1,\epsilon_2,\tau)$.
Similarly the contribution of anti-instantons localized to the south
pole is evaluated by an anti-topologically twisted theory, which leads
to Nekrasov's partition function with the argument $\bar\tau$.

\vskip2mm

So, our final result for the ellipsoid partition function is
\begin{eqnarray}
 Z &=&
 \int \td \hat a_0
 e^{-\frac{8\pi^2}{g_\text{YM}^2}\text{Tr}(\hat a_0^2)}
 |Z_\text{inst}|^2
 \prod_{\alpha\in\Delta_+}
 \Upsilon( i\hat a_0\cdot\alpha)
 \Upsilon(-i\hat a_0\cdot\alpha)
 \prod_{\rho\in R}\Upsilon(i\hat a_0\cdot\rho+\tfrac Q2)^{-1}.
\label{Ztot}
\end{eqnarray}

\paragraph{Wilson loops.}

The generalization of the above result to expectation values of
supersymmetric observables is straightforward. Of particular interest are
the Wilson loops. Supersymmetry requires the loops to be aligned with
the direction of $v$. When $\ell,\tilde\ell$ are incommensurable, there
are only two classes of closed loops. One of them winds along the
$\varphi$-direction and the other along $\chi$-direction, and they are
both labeled by $\rho$.
\begin{eqnarray}
 S^1_\varphi(\rho)&:& (x_0,x_1,x_2,x_3,x_4)=
 (r\cos\rho, \ell\sin\rho\cos\varphi,\ell\sin\rho\sin\varphi,0,0),
 \nonumber \\
 S^1_\chi(\rho)&:& (x_0,x_1,x_2,x_3,x_4)=
 (r\cos\rho, 0,0,\tilde\ell\sin\rho\cos\chi,\tilde\ell\sin\rho\sin\chi).
\end{eqnarray}
The corresponding supersymmetric Wilson loops are given by
\begin{eqnarray}
 W_\varphi(R)&\equiv&
 \text{Tr}_R\text{P}\exp i\int_{S^1_\varphi(\rho)}
  \td\varphi\Big(A_\varphi-2\ell(\phi\cos^2\tfrac\rho2
        +\bar\phi\sin^2\tfrac\rho2)\Big),
 \nonumber \\
 W_\chi(R)&\equiv&
 \text{Tr}_R\text{P}\exp i\int_{S^1_\chi(\rho)}
  \td\chi\Big(A_\chi-2\tilde\ell(\phi\cos^2\tfrac\rho2
        +\bar\phi\sin^2\tfrac\rho2)\Big).
\end{eqnarray}
At saddle points they take the classical values
\begin{eqnarray}
 W_\varphi(R)&=& 
 \text{Tr}_R\exp\left(-2\pi b\hat a_0\right),
 \nonumber \\
 W_\chi(R)&\equiv&
 \text{Tr}_R\exp\left(-2\pi b^{-1}\hat a_0\right).
\end{eqnarray}
The expectation values of Wilson loops can thus be computed by inserting
these expressions into the integral formula (\ref{Ztot}).

\section{Concluding Remarks}\label{sec:conclusion}

In this paper we have found an interesting deformation of the round
$S^4$ which supports SW theories with a rigid supersymmetry. In the
light of the AGT correspondence, our result for partition functions
should be related to the 2D Liouville or Toda correlators with the
coupling $b=(\ell/\tilde\ell)^{1/2}$. Let us quickly check this in the
simplest example of $SU(2)$ SQCD with four fundamental hypermultiplets,
which should correspond to Liouville four-point function on sphere. We
focus only on the one-loop part of the correspondence, since the other
parts, such as the coincidence between Nekrasov's instanton partition
function and Liouville conformal block, have already been extensively
investigated for general $b$.

For $SU(2)$ SQCD with four fundamental flavors, the saddle points are
labeled by a single parameter $p$, and the mass of the four
hypermultiplets $\mu_1,\cdots,\mu_4$ can be introduced via suitable
gauging of the $U(1)^4$ subgroup of the flavor group $SO(8)$.
The one-loop part of the partition function then reads
\begin{equation}
 Z_\text{1-loop} ~=~
 \frac{\Upsilon(2ip)\Upsilon(-2ip)}
      {\prod_{i=1}^4\Upsilon(\frac Q2+ip+i\mu_i)
                    \Upsilon(\frac Q2-ip+i\mu_i)}.
\end{equation}
To make correspondence with Liouville theory, we divide the four
hypermultiplets into two pairs, and associate each pair with the flavor
subgroup $SO(4)\simeq SU(2)\times SU(2)$. We thus get four copies of
$SU(2)$ flavor groups, and denote by $p_a$ the mass parameter associated
to the $a$-th $SU(2)$. The parameters $\mu_i$ and $p_a$ are related by
\cite{Gaiotto}
\begin{equation}
 \mu_1=p_1+p_2,\quad
 \mu_2=p_1-p_2,\quad
 \mu_3=p_3+p_4,\quad
 \mu_4=p_3-p_4.
\end{equation}
Under this identification $Z_\text{1-loop}$ agrees, up to some
$p$-independent factors, with the product of two Liouville 3-point
structure constants $C(p_1,p_2,p_3)$.
\begin{eqnarray}
Z_\text{1-loop} &\sim& C(p_1,p_2,p)C(p_1,p_2,-p)~=~
 C(p_1,p_2,p)C(p_1,p_2,p)R(p)^{-1}.
\end{eqnarray}
Here $R(p)\equiv\Upsilon(Q+2ip)/\Upsilon(Q-2ip)$ is the reflection
coefficient of Liouville primary operator with momentum
$\alpha=\frac Q2+ip$, and $C(p_1,p_2,p_3)$ is given by \cite{ZZ}
\begin{equation}
 C(p_1,p_2,p_3) =
 \frac{\text{const}\cdot\Upsilon(Q+2ip_1)\Upsilon(Q+2ip_2)\Upsilon(Q+2ip_3)}
      {\Upsilon(\frac Q2+ip_{1+2+3})
       \Upsilon(\frac Q2+ip_{1+2-3})
       \Upsilon(\frac Q2+ip_{1-2+3})
       \Upsilon(\frac Q2+ip_{1-2-3})}.
\end{equation}
Thus we found the agreement for general values of the coupling $b$.

\vskip2mm

We expect that, in comparison to ${\cal N}=1$ SUSY theories, the
theories with extended SUSY such as SW theories can be put on wider
class of backgrounds preserving supersymmetry, since the corresponding
off-shell supergravity multiplet contains more fields. It will be an
interesting problem to find and classify other 4D manifolds which have
solutions to our Killing spinor equation. Studying SW theories on such
manifolds may lead to yet another interesting generalization of the AGT
relation.

\vskip6mm

\paragraph{Acknowledgment.}

KH thanks Teruhiko Kawano, Takuya Okuda and Seiji Terashima for useful
and encouraging discussions, and Jaume Gomis for useful correspondences
regarding Fayet-Iliopoulos invariant. The work of NH is supported in
part by the JSPS Research Fellowships for Young Scientists.

\end{document}